\documentclass[lettersize,journal]{IEEEtran}
\usepackage{amsmath,amsfonts}
\usepackage{algorithmic}
\usepackage{algorithm}
\usepackage{array}
\usepackage[caption=false,font=normalsize,labelfont=sf,textfont=sf]{subfig}
\usepackage{textcomp}
\usepackage{stfloats}
\usepackage{url}
\usepackage{verbatim}
\usepackage{graphicx}
\usepackage{cite}
\usepackage{color}
\usepackage{url}
\usepackage{hyperref}
\usepackage{tabularx}
\usepackage{booktabs}
\newcommand{\orcid}[1]{\href{https://orcid.org/#1}{\includegraphics[width=8pt]{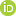}}}

\begin{document}

\title{

DurFlex-EVC: Duration-Flexible Emotional Voice Conversion Leveraging Discrete Representations without Text Alignment

}
\author{Hyung-Seok~Oh\orcid{0000-0001-7229-8123},
        Sang-Hoon~Lee\orcid{0000-0002-8925-4474},
        Deok-Hyeon~Cho\orcid{0009-0002-4673-9882},
        and~Seong-Whan~Lee\orcid{0000-0002-6249-4996},~\IEEEmembership{Fellow,~IEEE}
\thanks{This work was partly supported by Institute of Information \& Communications Technology Planning \& Evaluation (IITP) grant funded by the Korea government (MSIT) (No. RS-2019-II190079, Artificial Intelligence Graduate School Program (Korea University), No. RS-2021-II-212068, Artificial Intelligence Innovation Hub, IITP-2024-RS-2023-00255968, the Artificial Intelligence Convergence Innovation Human Resources Development and No. RS-2024-00336673, AI Technology for Interactive Communication of Language Impaired Individuals). \textit{(Corresponding author: Seong-Whan Lee.)}}
\thanks{H.-S. Oh, D.-H. Cho and S.-W. Lee are with the Department of Artificial
Intelligence, Korea University, 145, Anam-ro, Seongbuk-gu, Seoul 02841, Republic of Korea.\protect

E-mail: (hs\_oh@korea.ac.kr\; dh\_cho@korea.ac.kr\; sw.lee@korea.ac.kr). Sang-Hoon Lee is with the Department of Software and Computer Engineering and the Department of Artificial Intelligence, Ajou University, South Korea (e-mail: sanghoonlee@ajou.ac.kr)}
}

\markboth{}%
{}

\IEEEpubid{}

\maketitle

\begin{abstract}

Emotional voice conversion (EVC) involves modifying various acoustic characteristics, such as pitch and spectral envelope, to match a desired emotional state while preserving the speaker's identity. Existing EVC methods often rely on text transcriptions or time-alignment information and struggle to handle varying speech durations effectively. In this paper, we propose DurFlex-EVC, a duration-flexible EVC framework that operates without the need for text or alignment information. We introduce a unit aligner that models contextual information by aligning speech with discrete units representing content, eliminating the need for text or speech-text alignment. Additionally, we design a style autoencoder that effectively disentangles content and emotional style, allowing precise manipulation of the emotional characteristics of the speech. We further enhance emotional expressiveness through a hierarchical stylize encoder that applies the target emotional style at multiple hierarchical levels, refining the stylization process to improve the naturalness and expressiveness of the converted speech. Experimental results from subjective and objective evaluations demonstrate that our approach outperforms baseline models, effectively handling duration variability and enhancing emotional expressiveness in the converted speech.

\end{abstract}

\begin{IEEEkeywords}
emotional voice conversion, self-supervised representation, style disentanglement, duration control
\end{IEEEkeywords}

\section{Introduction}
\IEEEPARstart{E}{motional} voice conversion (EVC) involves modifying various acoustic characteristics of a voice, such as pitch and spectral envelope, to match a desired emotional state while preserving the speaker's identity\cite{10065433}.
EVC has gained prominence, particularly in the realm of voice-interactive technologies such as virtual assistants and internet of things (IoT) devices, improving the human-like and emotionally resonant aspects of digital interactions \cite{9829283, 8068274, busso2008iemocap, 9920692}.

In the context of EVC, a crucial objective is to preserve the speaker identity and content of the original speech while modifying only those speech attributes that convey emotion \cite{10065433, 7160715}. This necessitates an adjustment of prosody to align with the intended emotion. Prosody elements, including intonation, rhythm, and energy, play a critical role in both conveying and recognizing emotions in speech. Although the concept of controlling prosody for emotional conversion is intuitively appealing \cite{5871584}, refining each prosody component presents a significant challenge.

The field of EVC has been revolutionized by advances in deep learning \cite{ming16_interspeech, 9687906}. Initial approaches employed Gaussian mixture models \cite{aihara2012gmm} to convert spectral and prosody features, producing more expressive voices. Subsequent developments led to autoencoder-based methods \cite{gao19b_interspeech, zhou20d_interspeech, 9383526}, enabling learning in non-parallel data-driven EVC. VAE-based approaches \cite{cao20b_interspeech} and GAN-based frameworks \cite{9054579}—such as Cycle-GAN \cite{Zhu_2017_ICCV}, StarGAN \cite{Choi_2018_CVPR}, and VAE-GAN \cite{Bao_2017_ICCV}—represent further advances. 
However, these methods often overlook the importance of rhythm in expressing emotion, as they typically support emotional conversion with fixed durations.

Sequence-to-sequence (Seq2Seq) models, capable of implicitly modeling duration, have become a notable development \cite{8683865, 9053255}. These models often employ strategies, such as a two-stage learning approach integrating a text-to-speech (TTS) model, to improve stability \cite{zhou21b_interspeech, 9778970}.
However, Seq2Seq models face challenges like long-term dependency and repetition, necessitating parallel generation for efficiency and reliability. Parallel generation requires explicit duration modeling, which many voice conversion models \cite{9729483} achieve through phoneme duration derived from TTS alignments. 
Obtaining phoneme duration often requires encoder-decoder attention from pre-trained autoregressive TTS models\cite{NEURIPS2019_f63f65b5} or external forced alignment tools\cite{ren2021fastspeech}.

Recently, discrete speech units derived from self-supervised learning representations have shown promise in addressing parallel generation challenges \cite{polyak21_interspeech, kreuk-etal-2022-textless, 9746484, kim23k_interspeech}. 
Certain studies \cite{9746484, kim23k_interspeech} have proposed methods that model speech emotion conversion as a translation task, thereby enabling parallel audio generation. However, these methods still rely on autoregressive models for emotional translation, limiting their effectiveness in fully parallel generation.

In this paper, we propose a duration-flexible EVC framework, DurFlex-EVC, that eliminates the need for text or alignment information while supporting parallel generation. We use discrete speech units to model content and incorporate self-supervised representations to enrich acoustic details. Our model predicts unit sequences and their durations, which allows for flexible duration control. 

Our main contributions are as follows:
\begin{itemize} 
\item We propose DurFlex-EVC, a duration-flexible EVC framework that uses discrete speech units for content modeling, eliminating the need for text or external alignment information while supporting efficient parallel generation. 
\item We introduce a unit aligner that models contextual relationships between speech features and unit sequences, enabling effective duration control without relying on text-based alignment. 
\item We develop a style autoencoder that separates and reintroduces emotional styles in the input features, facilitating content-style disentanglement for emotional voice conversion.
\item We design a hierarchical stylize encoder that captures both global and local emotional patterns, thereby enhancing the expressiveness of emotional speech. 
\item We perform extensive subjective and objective evaluations, demonstrating the superior performance of our method compared to existing approaches. 
\end{itemize}

\section{Background}
\subsection{Exploring Self-Supervised Learning in Speech}
Self-supervised learning (SSL) is a machine learning paradigm in which models are trained on unlabeled datasets to generate meaningful representations. This approach is particularly advantageous in speech processing, where labeling data is labor-intensive and expensive. Models like wav2vec 2.0 \cite{NEURIPS2020_92d1e1eb} and vq-wav2vec \cite{Baevski2020vq-wav2vec} have demonstrated SSL's potential, with wav2vec 2.0 leveraging contrastive learning and vq-wav2vec introducing quantization techniques for discrete representations. Subsequent developments, such as XLS-R \cite{babu2021xls}, extend these ideas to cross-lingual applications, while models like Hidden-unit BERT (HuBERT) \cite{9585401} and ContentVec \cite{pmlr-v162-qian22b} focus on masked prediction and speaker disentanglement, respectively.

SSL representations have proven valuable across a wide range of speech-related tasks, including automatic speech recognition \cite{NEURIPS2020_92d1e1eb}, voice conversion \cite{lee23i_interspeech}, speaker verification \cite{9747814}, and speech emotion recognition \cite{10089511}. The increasing adoption of SSL techniques motivates further exploration into how these representations can be leveraged for more complex and resource-efficient speech applications, such as emotional voice conversion, where conventional supervised methods are often limited by data availability.

\subsection{Discrete Units in Speech Processing}
Discrete unit representations have emerged as a versatile tool in audio and speech processing \cite{9625818, dfossez2023high, yang2023uniaudio}. Techniques like SoundStream \cite{9625818} and EnCodec \cite{dfossez2023high} focus on high-fidelity audio compression, employing methods such as residual vector quantization (RVQ), while UniAudio \cite{yang2023uniaudio} serves as a general-purpose audio generation model. These neural codec-based methods, aimed at audio compression and restoration, utilize large codebooks and compact dimensions. Although these approaches excel in compression, they are less suited for tasks that require detailed modeling of linguistic and prosodic content.

In contrast, some methods emphasize encoding speech into semantic units to facilitate applications like speech synthesis \cite{kim23k_interspeech} and emotion conversion \cite{9585401}. For instance, techniques have been developed to decompose and reconstruct speech into discrete units for content, pitch, and speaker identity \cite{polyak21_interspeech}. To improve naturalness and intelligibility, soft speech units were introduced \cite{9746484} for enhanced content capture. Speech emotion conversion has also been framed as a language translation task \cite{kreuk-etal-2022-textless}, leveraging discrete representations of phonetic content, prosody, speaker, and emotion alongside neural vocoders for waveform generation. UnitSpeech \cite{kim23k_interspeech} has demonstrated effectiveness in personalized TTS and voice conversion, using self-supervised units to fine-tune a diffusion-based TTS model with minimal data, thus bypassing the need for retraining across tasks. This focus on discrete units offers a promising solution to the challenges of aligning speech content, offering the potential for more efficient and accurate modeling.

\subsection{Parallel Speech Generation}
Non-autoregressive speech synthesis models generate speech frames in parallel, significantly reducing inference time compared to autoregressive methods \cite{NEURIPS2019_f63f65b5,NEURIPS2020_5c3b99e8, pmlr-v139-popov21a}. Recent studies have demonstrated that parallel generation methods can offer superior performance and reliability over autoregressive approaches for applications like text-to-speech, voice conversion, and vocoders. In text-to-speech systems, parallel generation typically relies on precise alignments between text and speech. These alignments are often derived using pre-trained autoregressive teacher models \cite{NEURIPS2019_f63f65b5}, external aligners like the Montreal Forced Aligner (MFA) \cite{mcauliffe17_interspeech}, or algorithms such as the monotonic alignment search (MAS) \cite{NEURIPS2020_5c3b99e8}. Voice conversion methods often attempt to leverage the benefits of parallel generation.  While many approaches \cite{8639535, pmlr-v97-qian19c} have adopted parallel generation strategies, certain models remain autoregressive \cite{8683282} due to complexities in handling variable durations. To address these challenges, some solutions \cite{9413973, 9729483} have integrated variable-duration modeling with parallel generation. However, this often requires text-speech alignment similar to TTS frameworks. Additionally, vocoders have used techniques like GANs \cite{NEURIPS2020_c5d73680}, normalizing flows \cite{8683143}, and generative diffusion models \cite{kong2021diffwave} to efficiently produce high-quality waveforms. The success of parallel generation frameworks has inspired further exploration into adapting these techniques for emotional voice conversion, with the goal of achieving efficient and expressive speech synthesis without relying on traditional alignment methods.

\subsection{Duration Modeling in Speech Processing}
Duration modeling is crucial in speech synthesis, particularly in TTS systems, where discrepancies between character length and audio signals often occur \cite{NEURIPS2019_f63f65b5}. Early approaches used autoregressive models to implicitly handle duration, generating speech one frame at a time \cite{8461368, Li_Liu_Liu_Zhao_Liu_2019}. FastSpeech \cite{NEURIPS2019_f63f65b5} improved upon these methods by leveraging alignments from an autoregressive teacher model to explicitly model phoneme duration, thereby enabling parallel generation. Building on this, FastSpeech 2 \cite{ren2021fastspeech} introduced phoneme duration extraction using external forced alignment. Glow-TTS \cite{NEURIPS2020_5c3b99e8} further refined this approach by employing a monotonic alignment search to find optimal matches between text and latent representations. Duration modeling has also been explored in voice conversion, where seq2seq models are often used to manage variable durations \cite{8639647, 8683282}. The DCVC model \cite{9729483} utilized a phoneme-based information bottleneck for style transfer and speech speed control in voice conversion. Discrete speech units have also been leveraged in voice conversion for modeling duration by counting consecutive occurrences of each unit. This approach allows for effective duration control while bypassing the need for explicit text or phoneme alignment \cite{polyak21_interspeech, kim23k_interspeech}.
In EVC, seq2seq structures have become a popular choice for managing duration variations \cite{9778970, yang22t_interspeech}. \cite{kreuk-etal-2022-textless} leveraged discrete units for parallel emotional voice conversion but still incorporated seq2seq models for the translation of unit sequences, introducing challenges inherent to autoregressive architectures. Our work aims to address these limitations by employing discrete units to model duration, eliminating the need for text-speech alignment and enabling a more flexible parallel generation framework.

\begin{figure*}[!t]
    \centering
    \includegraphics[width=1.0\textwidth]{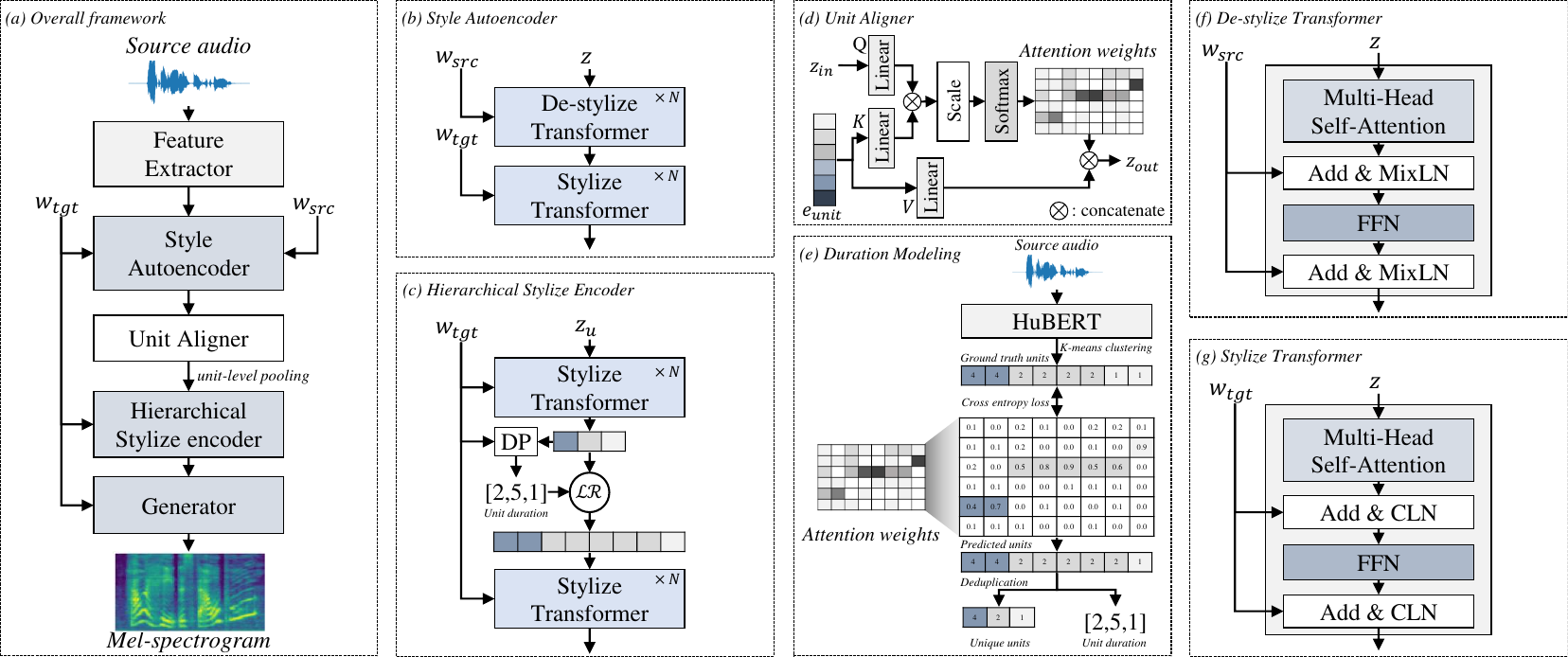}
    \caption{Overall framework of the proposed method. The feature extractor transforms the source audio into input features. These features are subsequently disentangled and reconditioned by the style autoencoder. The unit aligner is responsible for providing unit-level context and performing duration modeling. In addition, the hierarchical style encoder encodes features at both the unit and frame levels. Mel-spectrogram is subsequently produced by the generator. In this figure, "DP" represents the duration predictor, $\mathcal{LR}$ denotes the length regulator, while "Q", "K" and "V" represent the query, key and value of the cross-attention in the unit aligner, respectively. $\otimes$ denotes the concatenate operation. $w_{src}$ represents the source style vector and $w_{tgt}$ represents the target style vector. The style autoencoder disentangles the source style from the features and applies the target style, while the hierarchical stylize encoder and generator take the target style as a condition.}
\label{model}
\end{figure*}

\section{Proposed Method}
Our work introduces DurFlex-EVC, a duration-flexible and parallel generation framework for emotional voice conversion. We designed this system to overcome limitations of previous approaches, such as the reliance on seq2seq structures and the need for external text or alignment information. The key components of our model and their roles are as follows:
\begin{itemize} 
\item Feature extractor: Transforms raw audio waveforms into acoustic features. 
\item Style autoencoder: Designed to separate the emotional style from input features and apply the target emotional style, enabling content-style disentanglement without seq2seq dependencies. 
\item Unit aligner: Converts stylized features into unit-level content, capturing contextual information and handling duration modeling, which allows for efficient parallel context transformation. 
\item Hierarchical stylize encoder: Applies emotional styles at both unit and frame levels, ensuring that stylistic nuances are captured effectively. 
\item Diffusion-based generator: Produces high-quality Mel-spectrogram, which is then converted into speech by a pre-trained vocoder. 
\end{itemize}

\subsection{Overview}
Fig. \ref{model}a illustrates the overall architecture of DurFlex-EVC. The process begins with the feature extractor. It transforms raw audio waveforms into acoustic features. While various feature representations, such as Mel-spectrogram or SSL outputs, can be utilized, our approach employs the final layer representations from HuBERT \cite{9585401}. This choice provides continuous representations that retain comprehensive acoustic information, essential for accurately modeling emotional characteristics in speech. Next, the style autoencoder separates the emotional style from the input features and applies the desired target emotion. This disentanglement ensures that content and style are managed independently, allowing for precise emotional adaptation without relying on seq2seq structures. The unit aligner then processes the stylized features by aggregating contextual information at the frame level through a cross-attention module. This aggregated representation is compressed to the unit level, preparing it for further processing by the hierarchical stylize encoder. The hierarchical stylize encoder operates at both unit and frame levels, refining the emotional style to ensure consistency and expressiveness across different scales. Finally, the diffusion-based generator synthesizes a high-quality Mel-spectrogram from the output of the hierarchical stylize encoder and the style vector. This Mel-spectrogram is subsequently converted into a raw waveform using a pre-trained vocoder, completing the emotional voice conversion process.

\subsection{Style Autoencoder}
The feature extractor generates representations that encompass both content and style aspects of the speech. In this context, content refers to the linguistic information, while style includes emotional expression and speaker-specific characteristics. To model styles, we distinguish between speaker style and emotional style. The style autoencoder is designed to separate the source emotional style from the input features and apply a target emotional style, facilitating content-style disentanglement. Fig. \ref{model}b shows the structure of the style autoencoder, which consists of two primary components: the de-stylize transformer and the stylize transformer.

\textbf{Layer Normalization (LN)} is employed as a fundamental technique in both transformers, defined as: 
\begin{equation} \text{LN}(z) = \frac{z - \mu}{\sigma}, \end{equation} 
where $z$ represents the input vector, $\mu$ is the mean, and $\sigma$ is the standard deviation. LN ensures that the inputs to each layer have a consistent distribution, facilitating stable and efficient training.

\textbf{De-stylize Transformer}: This component removes the source emotional style from the input features, isolating the content. It employs mix-style layer normalization (MixLN) \cite{NEURIPS2022_4730d10b}, which blends the original style vector with a shuffled style vector to disrupt the association between features and the source style, as shown in Fig. \ref{model}f. The MixLN is defined as:
\begin{equation}
    \text{MixLN}(z,w) = \gamma_{mix}(w) \times \text{LN}(z) + \beta_{mix}(w),
\end{equation}
where $\gamma_{mix}(w) = \lambda\gamma(w) + (1 - \lambda)\gamma(\tilde{w})$ and $\beta_{mix}(w) = \lambda\beta(w) + (1 - \lambda)\beta(\tilde{w})$. 
$w$ is the original style vector, $\tilde{w}$ is the shuffled style vector, and $\lambda$ is a parameter sampled from a Beta distribution, $\text{Beta}(\alpha, \alpha)$. This process helps in disentangling style-independent content by preventing the model from encoding style-specific features.

\textbf{Stylize Transformer}: This component applies the target emotional style to the content features. It employs conditional layer normalization (CLN) \cite{chen2021adaspeech}, which adjusts the normalized input based on the target style vector $w$, as illustrated in Fig. \ref{model}g. The CLN is defined as:
\begin{equation}
    \text{CLN}(z,w) = \gamma(w) \times \text{LN}(z) + \beta(w),
\end{equation}
where $\gamma(w)$ and $\beta(w)$ are adaptive parameters derived from the style vector $w$. CLN allows the model to incorporate the desired emotional style into the content features seamlessly.

The style autoencoder consists of $N$ de-stylize transformers and $N$ stylize transformers. We construct the source style vector $w_{src}$ by combining the speaker vector $s_{src}$ and the emotion vector $e_{src}$, and the target style vector $w_{tgt}$ by combining $s_{src}$ with the desired emotion vector $e_{tgt}$:
\begin{align}
    w_{src} &= s_{src} + e_{src}, \\
    w_{tgt} &= s_{src} + e_{tgt}.
\end{align}

Both the speaker vector $s_*$ and the emotion vector $e_{*}$ are obtained from the embedding look-up table. The de-stylize transformer uses $w_{src}$ to disentangle the source emotional style from the input features, while the stylize transformer applies $w_{tgt}$ to encode the target emotional style into the features. This design enables the model to adapt emotional styles effectively without relying on seq2seq structures, thereby supporting parallel processing and flexible duration modeling.

\begin{figure}[!t]
\centering
\subfloat[unit-level pooling]{\includegraphics[width=0.45\columnwidth]{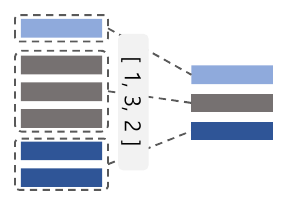}
\label{unit_level_pooling}}\hfill
\subfloat[frame-level scaling]{\includegraphics[width=0.45\columnwidth]{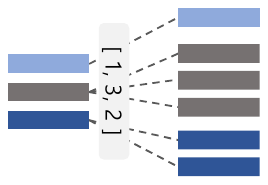}
\label{frame_level_scaling}}
\caption{Unit-level pooling and frame-level scaling. (a) Latent is pooled on average based on unit durations, and (b) Latent is expanded by being duplicated a number of times corresponding to the duration count.}
\end{figure}

\subsection{Unit Aligner}
The unit aligner is responsible for predicting unit sequences by modeling the semantic context of the input speech. As illustrated in Fig. \ref{model}d, the unit aligner leverages a cross-attention mechanism combined with learnable embeddings to generate attention weights that guide unit prediction.

We use the output of the style autoencoder as a query ($Q$) and introduce learnable embeddings $e_{unit}$ as keys ($K$) and values ($V$) to cross-attention. The attention weights $\mathcal{A}_{unit}$ are computed as follows:

\begin{equation}
\mathcal{A}_{unit} = \text{softmax}\left(\frac{QK^T}{\sqrt{d}}\right),
\end{equation}
where $d$ is the dimension of $Q$ and $K$. Subsequently, these weights $\mathcal{A}_{unit}$ are integrated with the value matrix $V$ to produce the attention output $z_{attn}$.

\begin{equation}
z_{attn} = \mathcal{A}_{unit} \cdot V.
\end{equation}

Fig. \ref{model}e shows the process of learning attention weights to predict units and the modeling of the duration of the predicted units. We introduce an additional loss term to guide the attention module to learn semantic information. This approach implies a direct classification task by aligning the attention weights $\mathcal{A}_{unit}$ with the target unit sequence $y$:

\begin{equation}
\mathcal{L}_{unit} = -\frac{1}{L}\sum_{i=1}^{L}\sum_{j=1}^{C} y_{i,j} \log(\mathcal{A}_{unit}^{i,j}),
\end{equation}
where $L$ is the length of the unit sequence and $C$ is the number of unit classes, $ \mathcal{A}_{unit}^{i,j} $ represents the predicted probability of the $i$-th element for class $j$ from the attention weights $\mathcal{A}_{unit}$, and $y_{i,j}$ is the one-hot encoded class label for the $i$-th unit for class $j$. 
We adopt the HuBERT unit as our target unit. This is an intended design feature that ensures that the context reflected in the unit aligner is consistent with the target style, rather than based on the context of the input speech.
During the emotional voice conversion process, the style autoencoder first removes the source emotion from the input features and applies the target emotion. 
The unit aligner then processes these stylized features to predict the corresponding unit sequences. This design enables the model to transform context efficiently. It also supports parallel processing without relying on seq2seq structures.

\subsection{Duration Modeling}
Once the unit aligner has predicted the unit sequence, the duration modeling component determines the duration of each unit based on consecutive identical units. This allows the model to adjust the generated speech length to align with the target emotional style. The unit sequence $\hat{y}$ is obtained by selecting the unit with the highest probability for each frame:

\begin{equation}
\hat{y}_i = \arg\max(\mathcal{A}_{unit}^{i}),
\end{equation}
where $\hat{y}^i$ is the $i$-th predicted unit and $\mathcal{A}_{unit}^{i}$ represents the attention weights for the $i$-th frame. To extract a distinct sequence of units and their consecutive counts, a deduplication operation is applied:
\begin{equation}
\hat{y}_{uniq}, n_{count} = \text{dedup}(\hat{y}),
\end{equation}
where $\hat{y}_{uniq}$ contains the sequence of unique units, and $n_{count}$ represents the number of consecutive occurrences for each unit. For example, given an input sequence $\hat{y}=[4,4,2,2,2,2,1,1]$, the deduplication results in $\hat{y}_{uniq}=[4,2,1]$ and $n_{count}=[2,4,2]$. This indicates two instances of unit $4$, followed by four instances of unit $2$ and two instances of unit $1$. 

The duration predictor is trained using $n_{count}$ as the target durations. This predictor estimates the number of consecutive units based on the unit-level representations generated by the unit aligner. Fig. \ref{unit_level_pooling} explains unit-level pooling. The output of the unit aligner, $z_{attn}$, is averaged based on the duration of the unit and results in downsampling the sequence length for alignment at the unit level.

\begin{equation}
z_{u} = \text{unit-level-pooling}(z_{attn}, n_{count}),
\end{equation}
where $z_{u}$ is the latent downsampled at the unit level. 
For example, given that $z_{attn}=[0.2,0.2,0.1,0.4,0.5,0.2,0.3,0.5]$ and $n_{count}=[2,4,2]$, the result of unit-level pooling is $z_{u}=[0.2, 0.3, 0.4]$. 

\subsection{Hierarchical Stylize Encoder}
The hierarchical stylize encoder \cite{10095515} functions at two levels: the unit level and the frame level. It consists of two components: the unit-level stylize transformer (UST) and the frame-level stylize transformer (FST). UST processes $z_{u}$ into a $z_{us}$, denoted as $z_{us} = \text{UST}(z_{u}, w_{tgt})$, focusing on unit-specific features. This refined variable $z_{u}$ is scaled to the frame level through a length regulator $\mathcal{LR}$, depicted in Fig. \ref{frame_level_scaling}.

\begin{equation}
z_{f} = \mathcal{LR}(z_{us}, n_{count}),
\end{equation}
where $z_{f}$ represents the latent variable at the frame level. For example, if $z_{us}=[0.1, 0.2, 0.5]$ and $n_{count}=[2,5,1]$, then $z_{f}$ becomes $[0.1,0.1,0.2,0.2,0.2,0.2,0.2,0.5]$. The FST further refines the frame-level features $z_{f}$ to $z_{fs}$, expressed as $z_{fs} = \text{FST}(z_{f}, w_{tgt})$.
This final output $z_{fs}$ is subsequently used as input for the Mel-spectrogram generator.

The duration predictor takes $z_{us}$ and $w_{tgt}$ as input and is trained to predict the unit-level duration $n_{count}$. For emotion-based duration dynamics, we introduce the flow-based stochastic duration predictor proposed in \cite{pmlr-v139-kim21f} to add duration uncertainty. The duration predictor training objective $\mathcal{L}_{dur}$ follows a negative variational lower bound. 

\subsection{Diffusion-Based Mel-Spectrogram Generator}
We use a diffusion framework based on stochastic differential equations (SDE) to generate high-quality speech with expressive emotions. The diffusion-based model gradually transforms the Mel-spectrogram into Gaussian noise in a forward process and generates samples from the noise in a reverse process. We adopt the standard normal distribution as the prior distribution, as in \cite{kim23k_interspeech}. The model is trained to minimize the mean square error (MSE) loss $\mathcal{L}_{diff}$ between the ground truth noise and the estimated noise. For score estimation, our model incorporates a network denoted by $s_\theta$ based on the U-net architecture with linear attention used in Grad-TTS \cite{pmlr-v139-popov21a}.

\subsection{Training Objective}
Consequently, the model is trained using the following loss function:
\begin{equation}
    \mathcal{L}_{total} = \lambda_{diff}\mathcal{L}_{diff} + \lambda_{unit}\mathcal{L}_{unit} + \lambda_{dur}\mathcal{L}_{dur},
\end{equation}
where $\lambda_{diff}$, $\lambda_{unit}$, and $\lambda_{dur}$ are the loss weights, which we set to 1.0, 0.1, and 0.1, respectively. 

\subsection{Emotion Voice Conversion Process}
The process of converting the emotion in the input speech to the target emotion is as follows. 
\begin{enumerate}
    \item The input waveform is converted into input features by the feature extractor. 
    \item The features are de-stylized from the source style vector and stylized to the target style vector by the style autoencoder. 
    The source style vector is obtained from the source emotion vector and the speaker vector. 
    The target style vector is obtained from the target emotion vector and the speaker vector. 
    The source style is disentangled by the MixLN of the style autoencoder, and the target style is applied by the CLN. 
    \item Unit-level features according to the target style are obtained with the unit aligner. 
    \item The hierarchical stylize encoder adapts the target style to the features at the unit-level and the frame-level.
    \item The diffusion-based generator produces a Mel-spectrogram conditioned on the feature and target style vector. 
    \item The waveform is synthesized by pre-trained vocoder. 
\end{enumerate}

\begin{figure}[!t]
\centering
\subfloat[Speaker]{\includegraphics[width=0.5\linewidth]{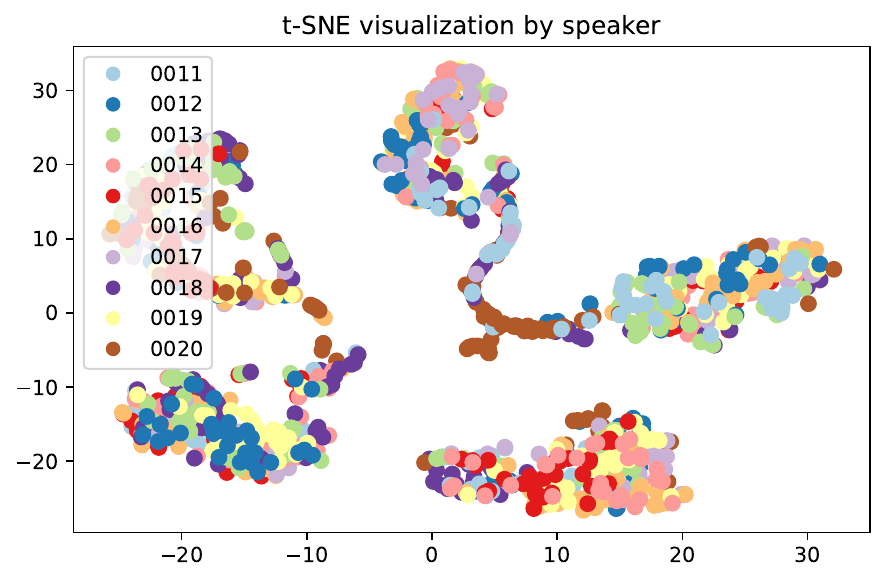}}
\subfloat[Emotion]{\includegraphics[width=0.5\linewidth]{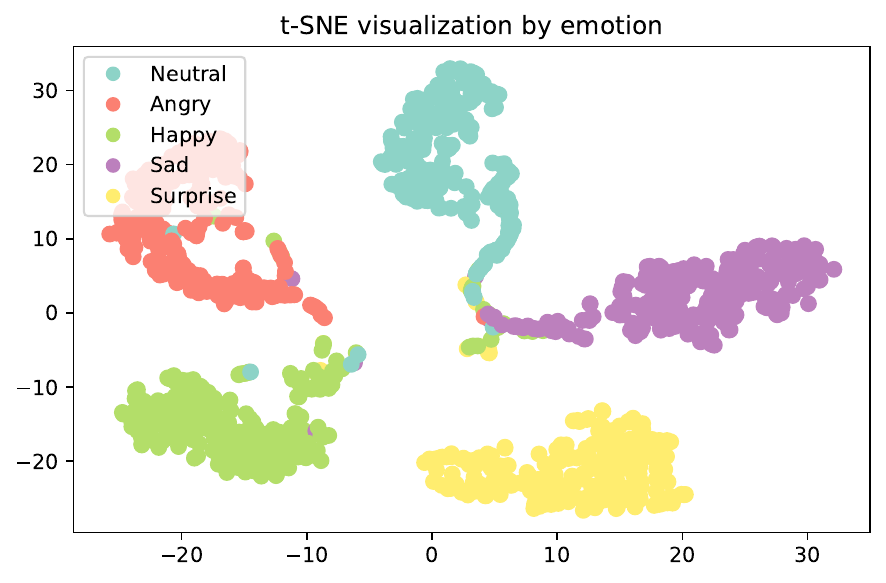}}
\caption{Visualize t-SNE of emotion2vec features for speaker and emotion.}
\label{fig:tsne}
\end{figure}

\section{Experiments}

\subsection{Experimental Setup}
We conducted experiments using the emotional speech dataset (ESD)\footnote{\url{https://github.com/HLTSingapore/Emotional-Speech-Data}}\cite{ZHOU20221}, which contains 350 parallel utterances spoken by 10 native Mandarin speakers and 10 English speakers with 5 emotional states (neutral, happy, angry, sad, and surprise). Following the data partitioning guidelines provided by ESD, we constructed the training set with 300 samples per emotion per speaker, for a total of 15,000 samples. The validation set included 20 samples for each emotion per speaker, totaling 1,000 samples, and the test set comprised 30 samples for each emotion per speaker, totaling 1,500 samples. Audio was sampled at 16,000 Hz and transformed into an 80-bin Mel-spectrogram using a short-time Fourier transform (STFT) with a window length of 1,024 and a hop size of 256. 

The experiments included transformations between all possible emotional states, not just limited from neutral to other states. This approach was designed to cover all possible emotional state conversions, ensuring a comprehensive assessment of the model's performance. For subjective evaluation, 10 sentences were randomly selected for each of the 5 emotions. These sentences were then adapted to reflect each of the four other emotional states, resulting in a total of 200 samples ($10 \times 5 \times 4 = 200$). For objective evaluation, each of the 1500 test samples was transformed into the four other emotional states, resulting in a total of 6000 samples ($1500 \times 4 = 6000$). This setup provided a thorough assessment of the model's performance across all possible emotional state conversions.

\subsection{Implementation Details}
In our experimental setup, we configured both the de-stylize and stylize transformers with specific parameters: the hidden dimension, kernel size, number of heads, FFN kernel size, and feed forward network (FFN) hidden size were set to 256, 5, 2, 9, and 1024, respectively. The $\alpha$ parameter of the Beta distribution for MixLN was fixed at 0.1. All transformers used in our model were organized into $N$ layers, with $N$ set to 4. The unit aligner featured multi-head attention with 16 heads. We set $T = 1$, $\beta_t = \beta_0 + (\beta_1-\beta_0)t$, $\beta_0 = 0.05$, and $\beta_1 = 20$ as noise schedules. The U-Net in our model was set to downsample four times and had a hidden dimension of 128. We evaluated our model with two different time steps: 4 and 100. The duration predictor, which comprises residual blocks using dilated and depth-separable convolution, was structured in four layers. To address the resolution disparity between the HuBERT unit and the Mel-spectrogram, we expanded the hidden representation with a length regulator and employed linear interpolation for upsampling. In training the generator, we utilized random segments, setting the segment size to 32 frames of the Mel-spectrogram. The AdamW optimizer was used, with a learning rate of $1\times 10^{-4}$. We set the batch size to 16 and the number of training steps to 500K.
We trained the vocoder using the official BigVGAN\footnote{\url{https://github.com/NVIDIA/BigVGAN}}\cite{lee2023bigvgan} implementation, incorporating the LibriTTS \cite{zen19_interspeech}, VCTK \footnote{\url{https://datashare.ed.ac.uk/handle/10283/2651}}, and ESD datasets. 
All experiments were conducted on an Intel® Xeon® Gold 6148 CPU @ 2.40 GHz  and a single NVIDIA RTX A6000 GPU. For broader accessibility, the code\footnote{\url{https://github.com/hs-oh-prml/DurFlexEVC}} and a demo\footnote{\url{https://prml-lab-speech-team.github.io/durflex/}} of our proposed method are available online.

\subsection{Evaluation}
\subsubsection{Subjective Metrics}
We conducted subjective evaluations using Amazon Mechanical Turk (mTurk). Our analysis included the mean opinion score (MOS) for naturalness (nMOS) and speaker similarity (sMOS), using a 9-point scale ranging from 1 to 5, with increments of 0.5 units. The results are presented with a confidence interval (CI) of 95\%.
Furthermore, we use the emotion mean opinion classification (eMOC) as suggested in \cite{kreuk-etal-2022-textless}. 

\begin{table}[!t]
\centering
\caption{Performance of the pre-trained evaluator model on test sets}
\label{tab:gt}
\resizebox{1.0\linewidth}{!}{
\begin{tabular}{l|ccccccc}
\toprule
        Model & UTMOS & PER & CER & WER & ECA & EECS & SECS \\
        \midrule
        GT              & 3.60 & 11.64 & 3.06 & 12.09 & 96.33 & 0.93 & 0.81 \\
        GT (vocoded)    & 3.58 & 11.73 & 3.14 & 12.45 & 94.13 & 0.91 & 0.81 \\      
\bottomrule
\end{tabular}    
}
\end{table}

\begin{figure}[!t]
\centering
\subfloat[StarGAN-EVC]{\includegraphics[width=0.5\linewidth]{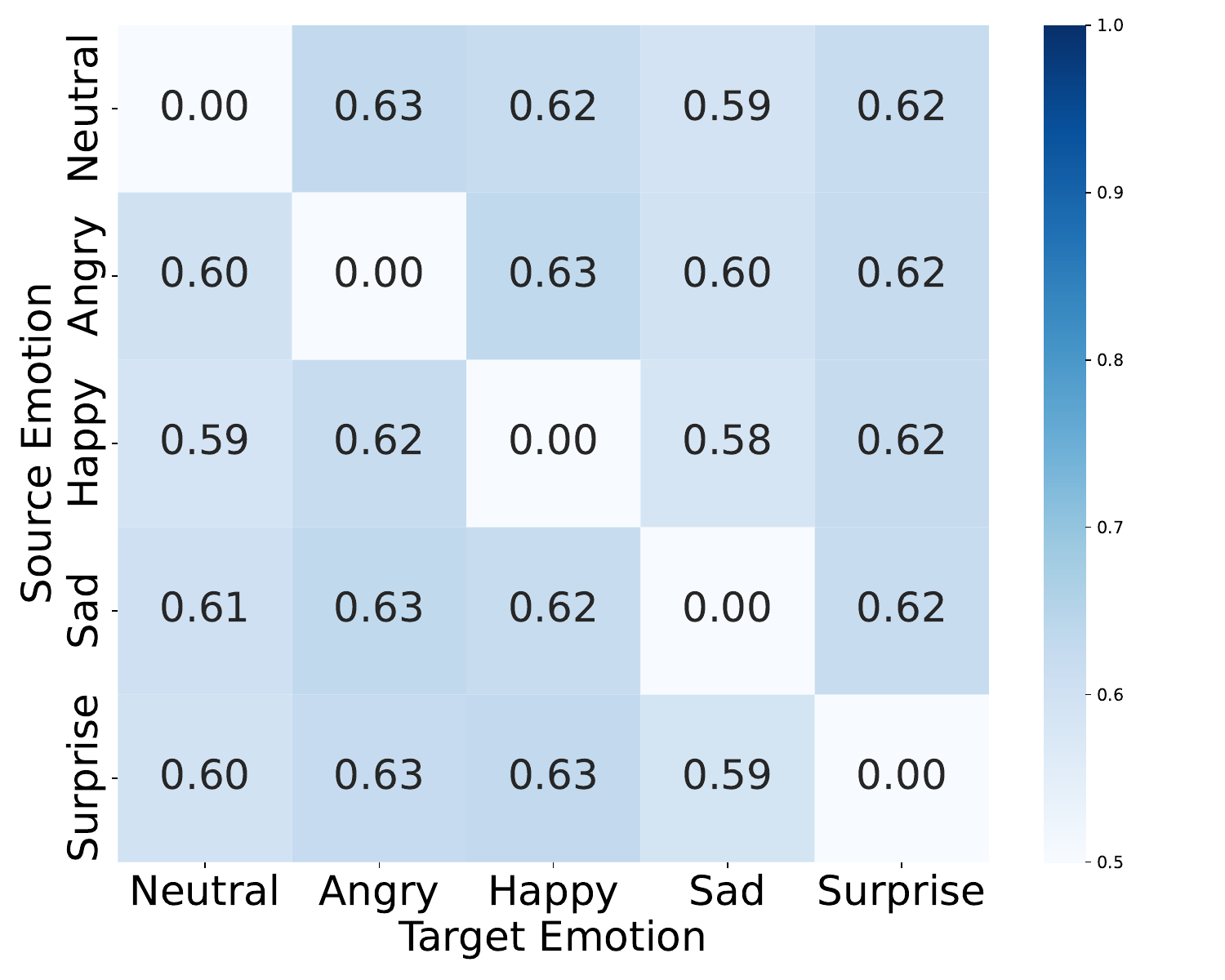}}
\subfloat[Seq2seq-EVC]{\includegraphics[width=0.5\linewidth]{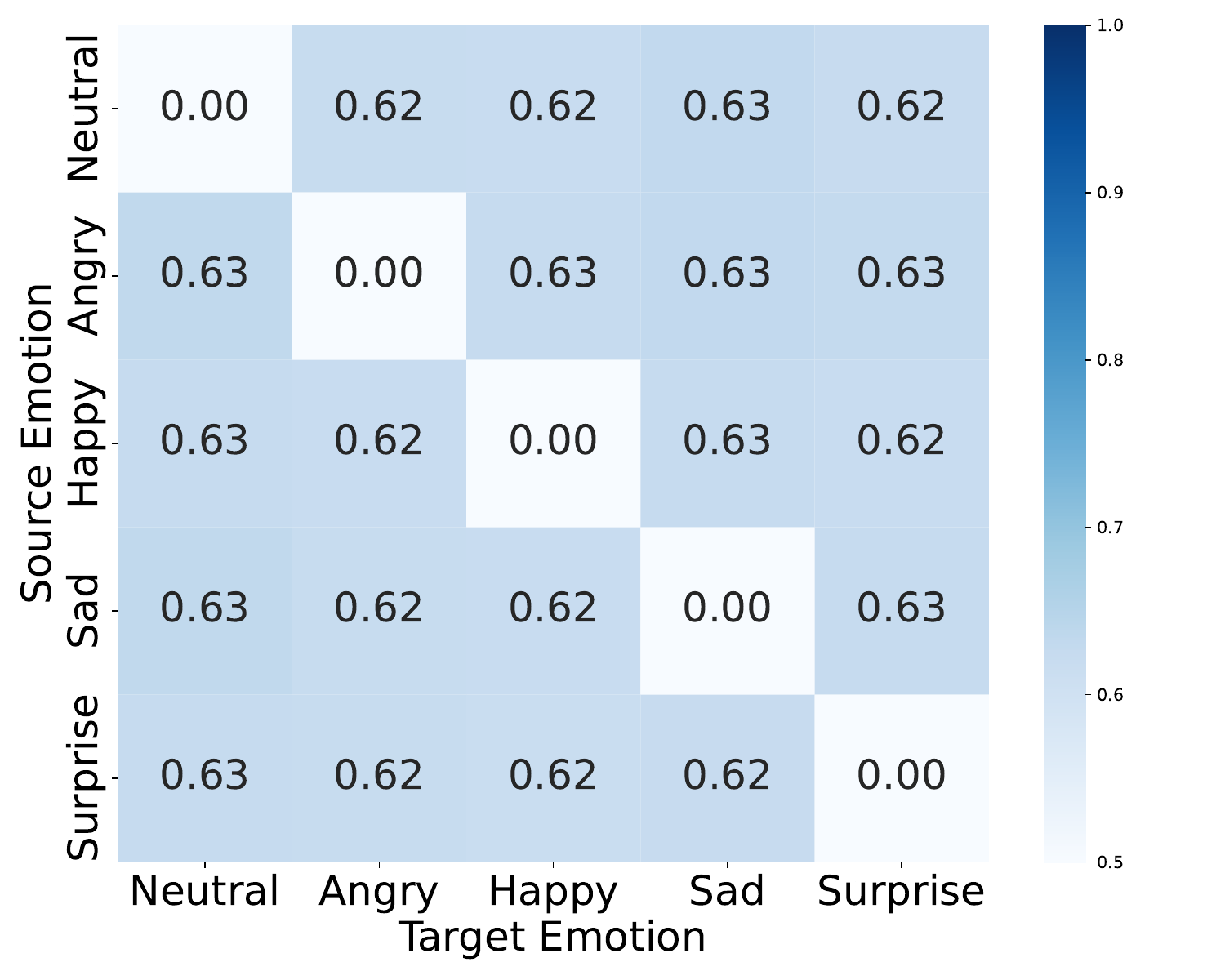}}

\subfloat[Emovox]{\includegraphics[width=0.5\linewidth]{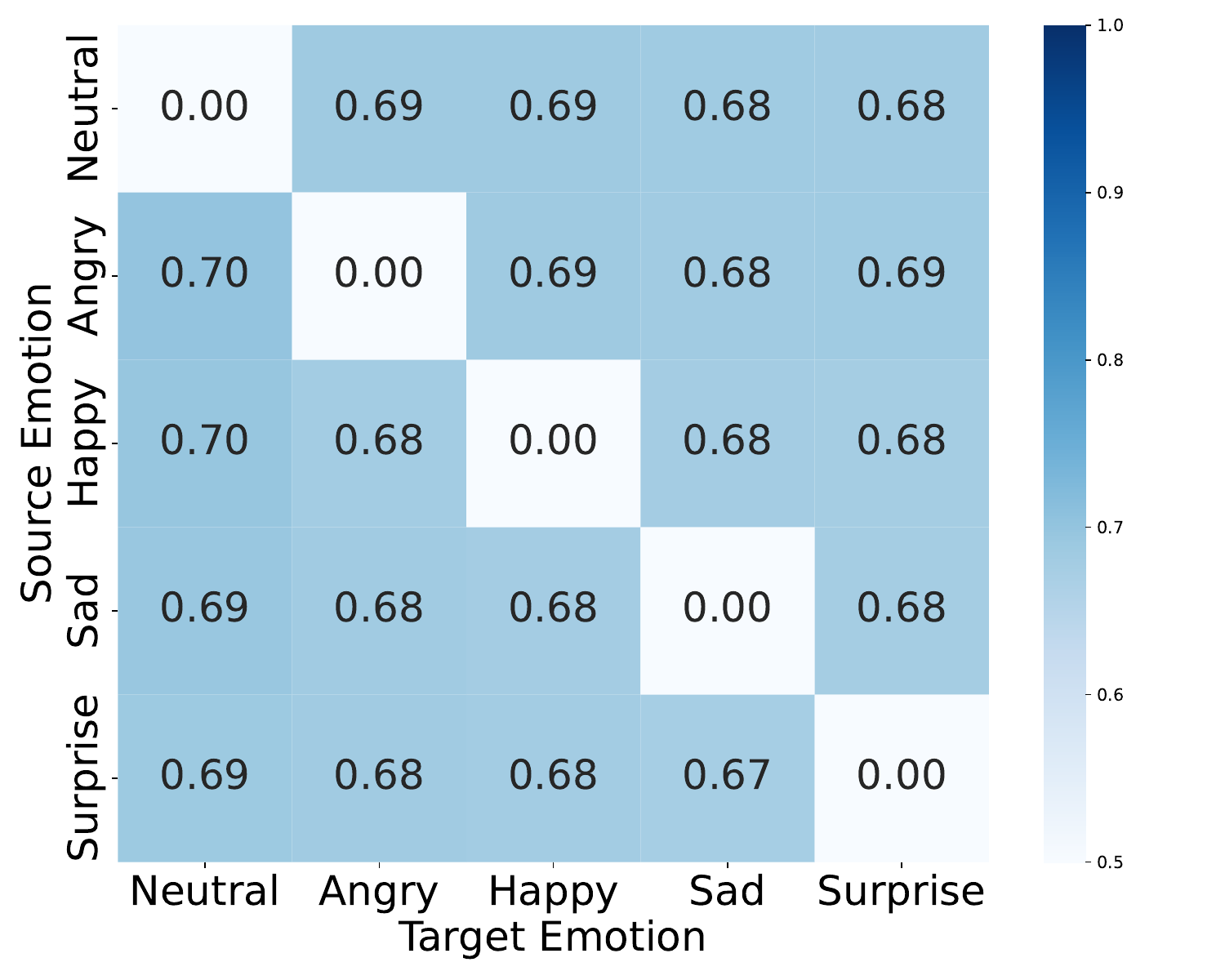}}
\subfloat[Mixed-Emotions]{\includegraphics[width=0.5\linewidth]{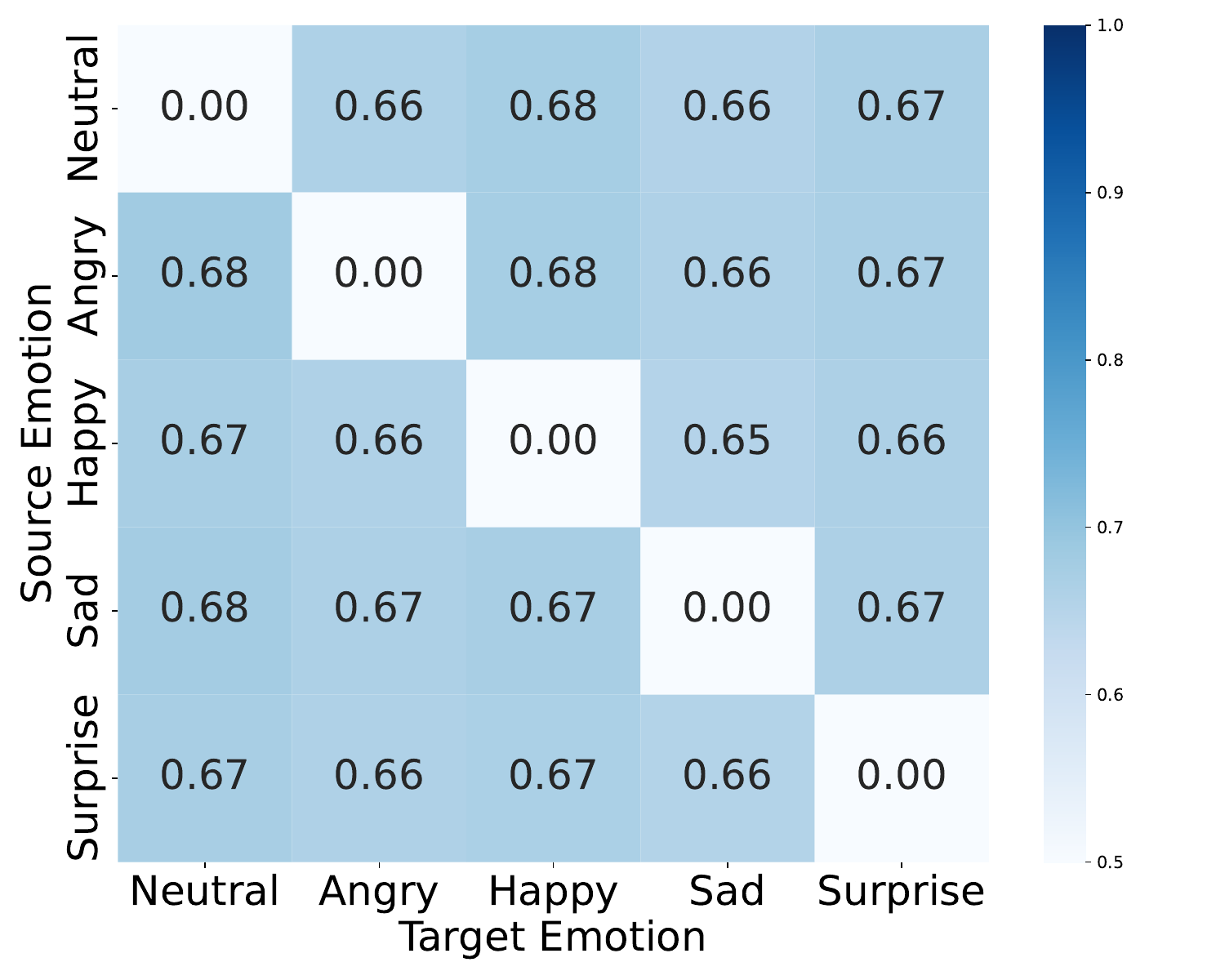}}

\subfloat[Textless-EVC]{\includegraphics[width=0.5\linewidth]{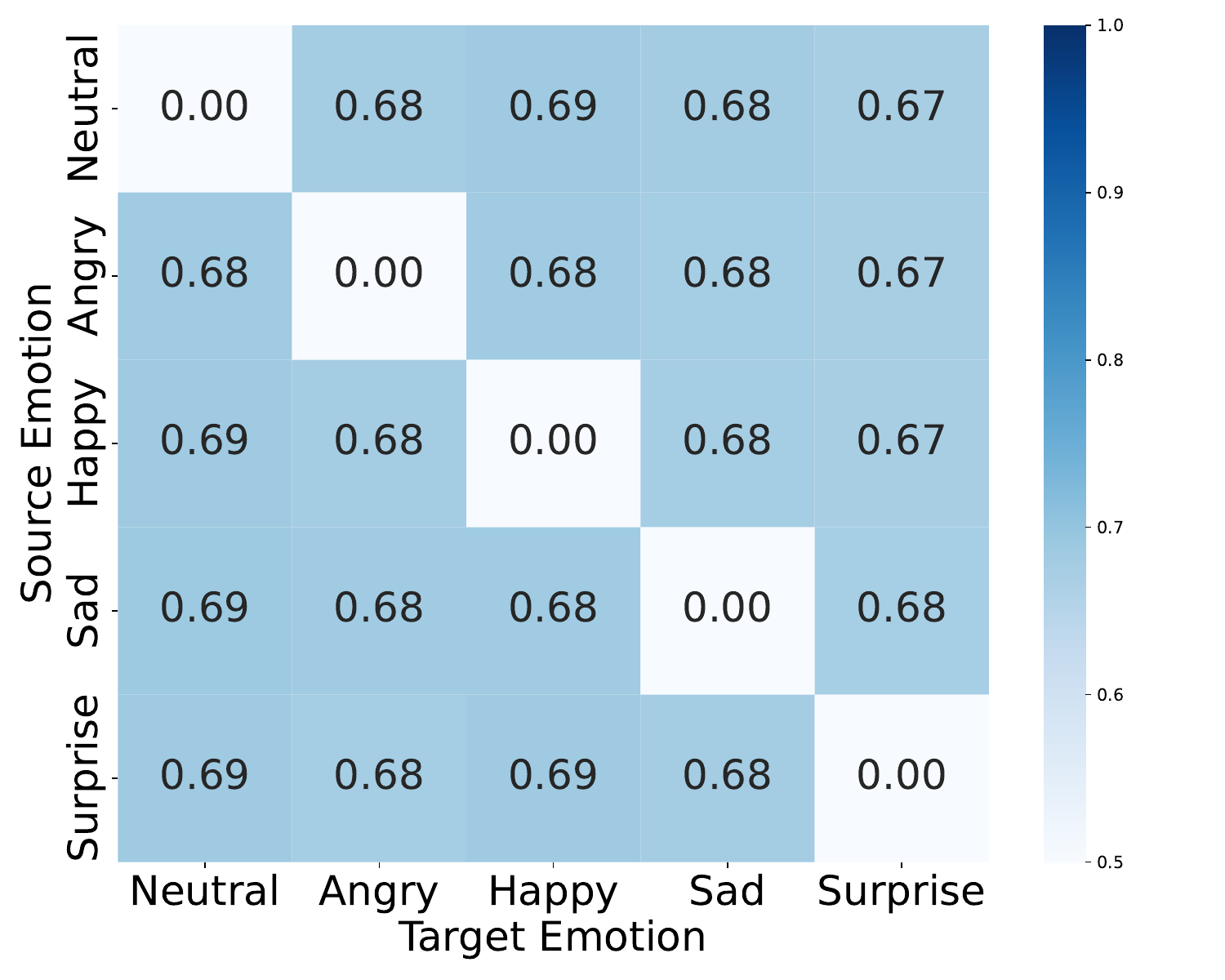}}
\subfloat[DurFlex-EVC]{\includegraphics[width=0.5\linewidth]{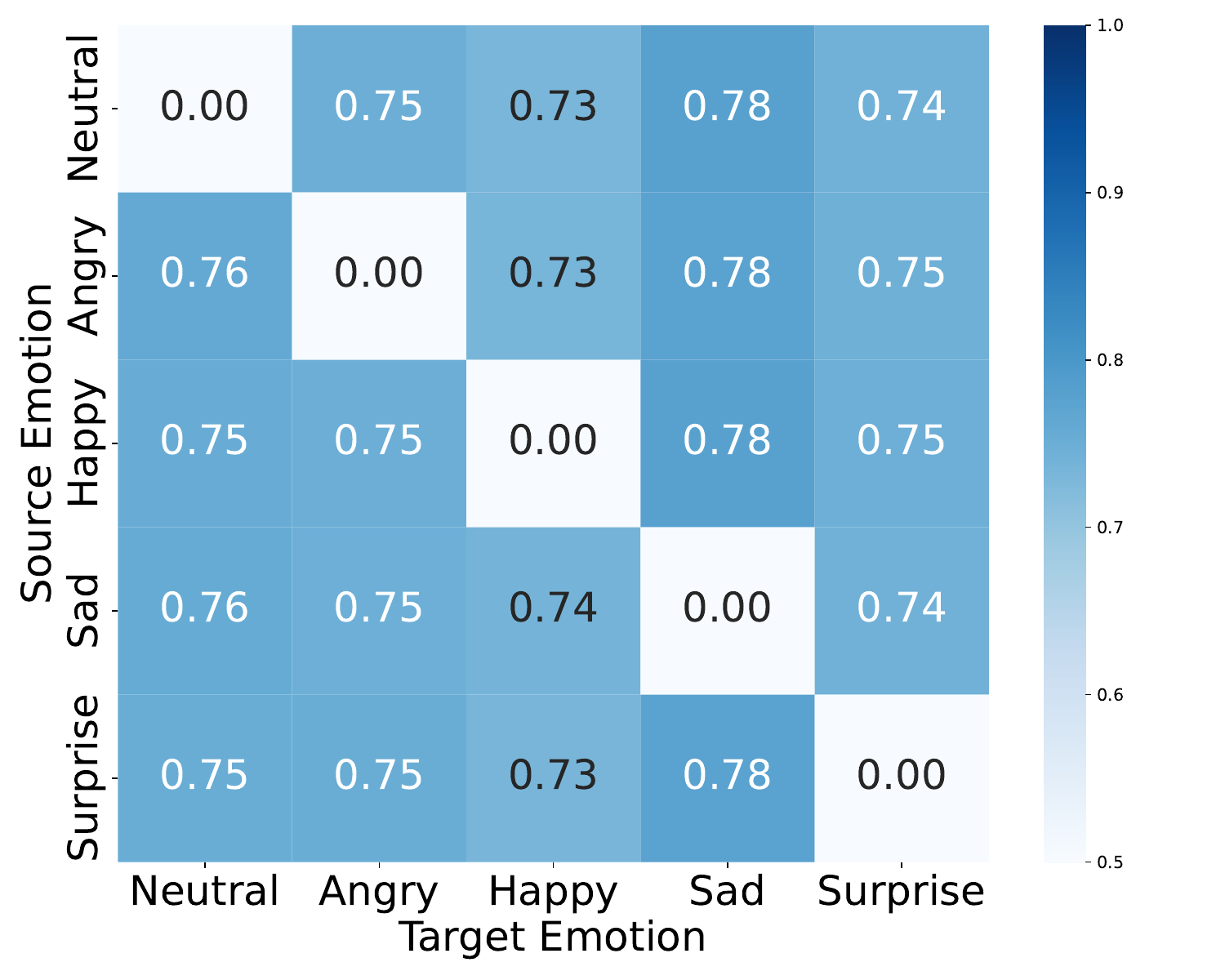}}
\caption{Comparison of the SECS scores of the comparison models for all combinations of emotion conversion.}
\label{fig:secs_comparison}
\end{figure}

\begin{figure}[!t]
\centering
\subfloat[StarGAN-EVC]{\includegraphics[width=0.5\linewidth]{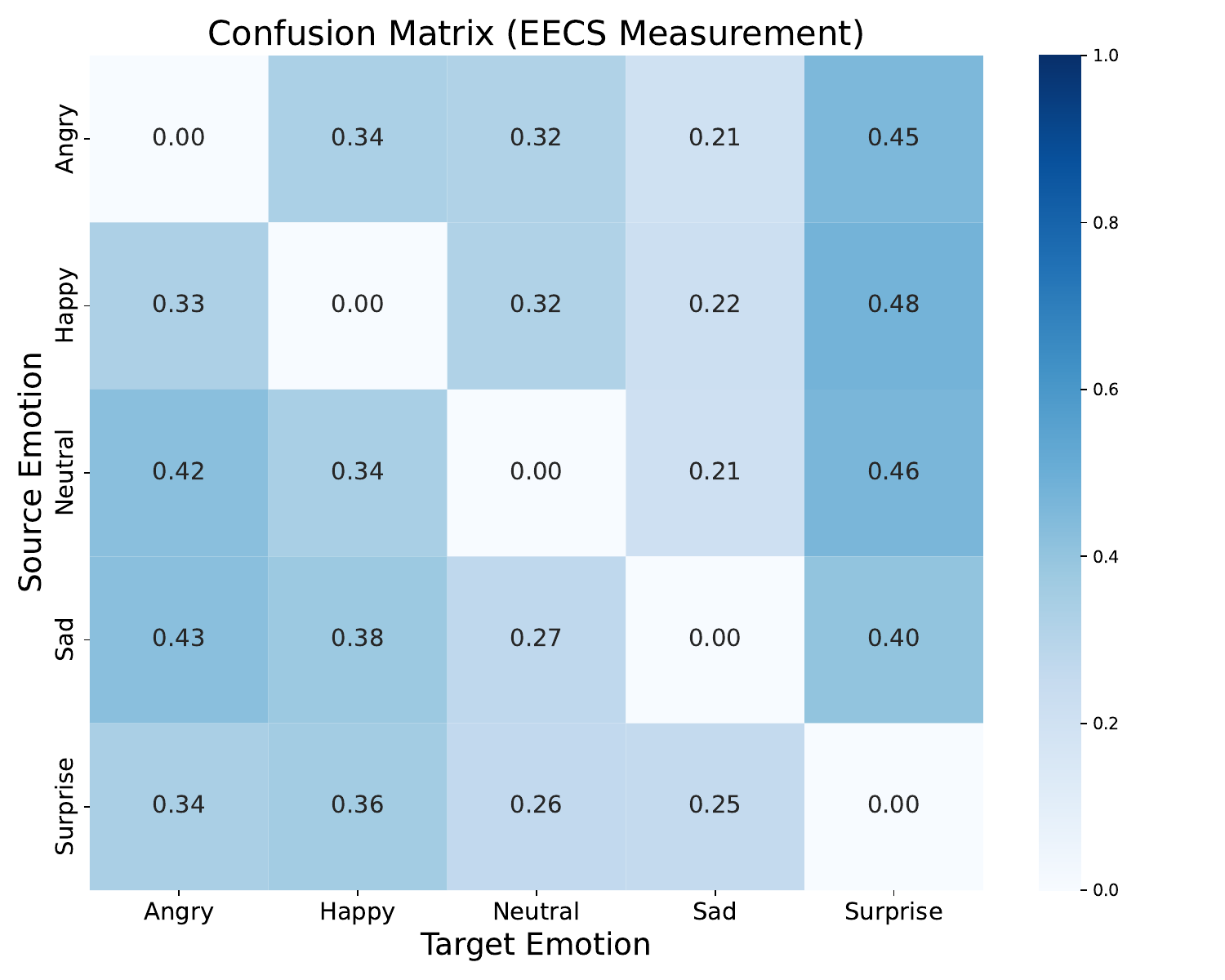}}
\subfloat[Seq2seq-EVC]{\includegraphics[width=0.5\linewidth]{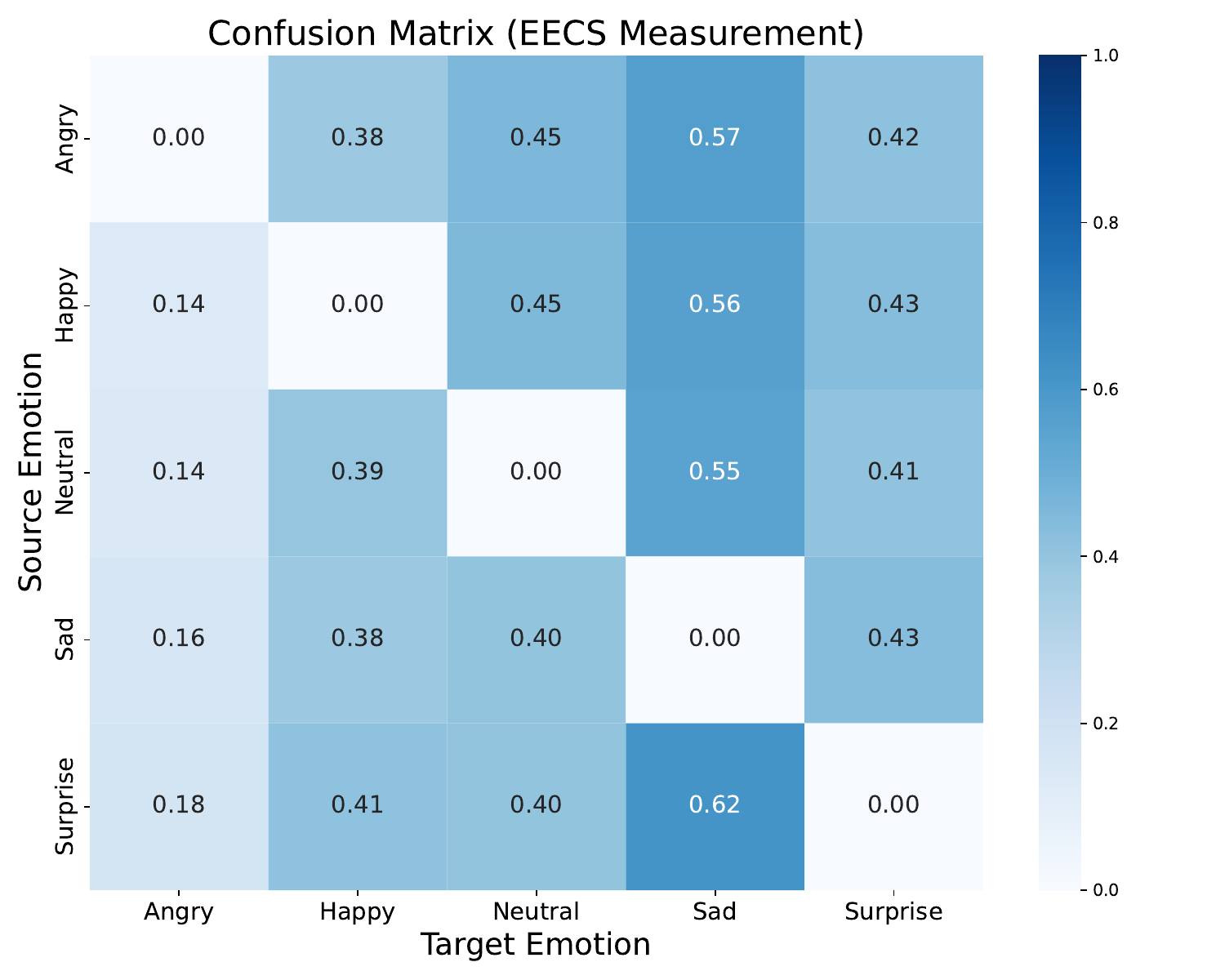}}

\subfloat[Emovox]{\includegraphics[width=0.5\linewidth]{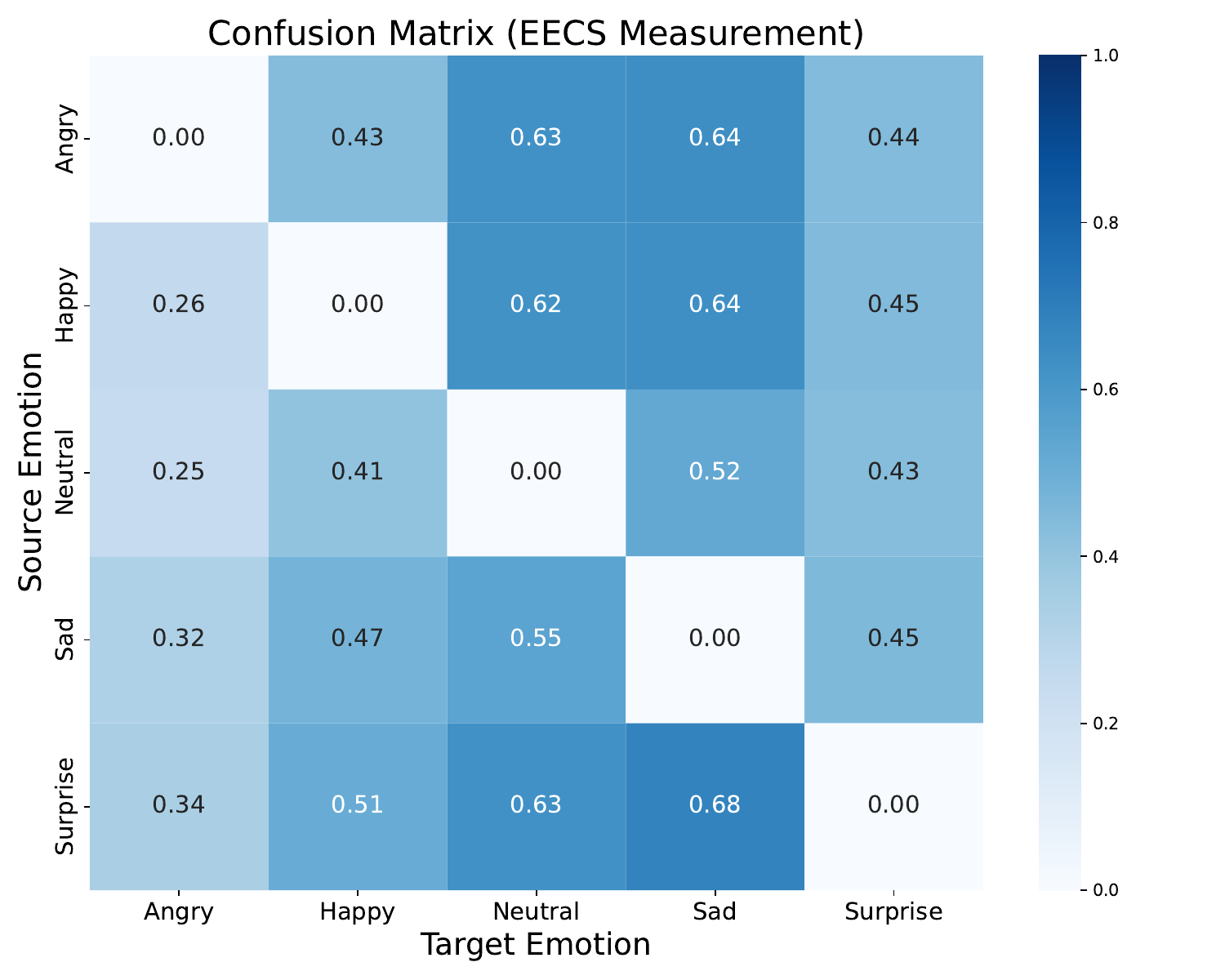}}
\subfloat[Mixed-Emotions]{\includegraphics[width=0.5\linewidth]{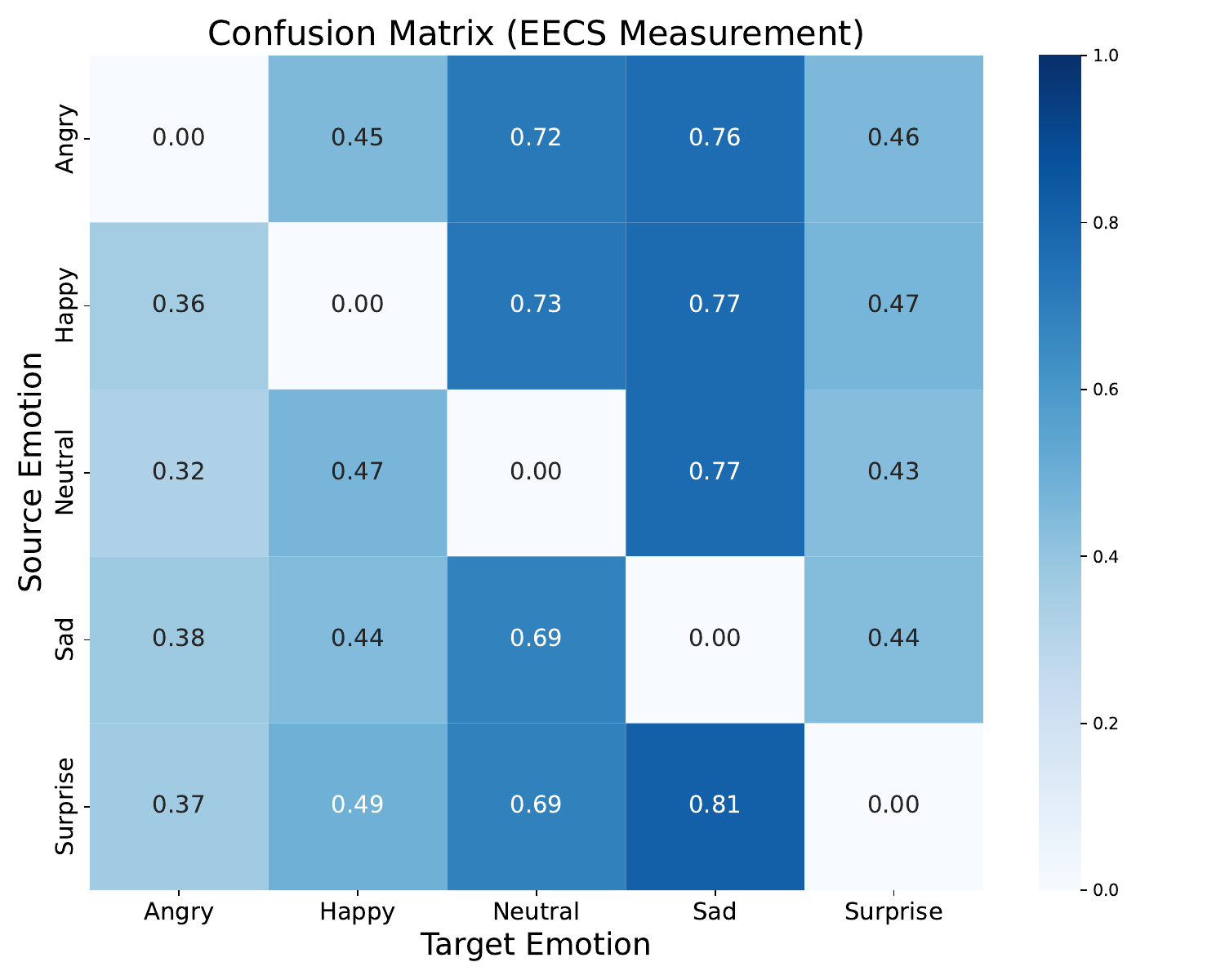}}

\subfloat[Textless-EVC]{\includegraphics[width=0.5\linewidth]{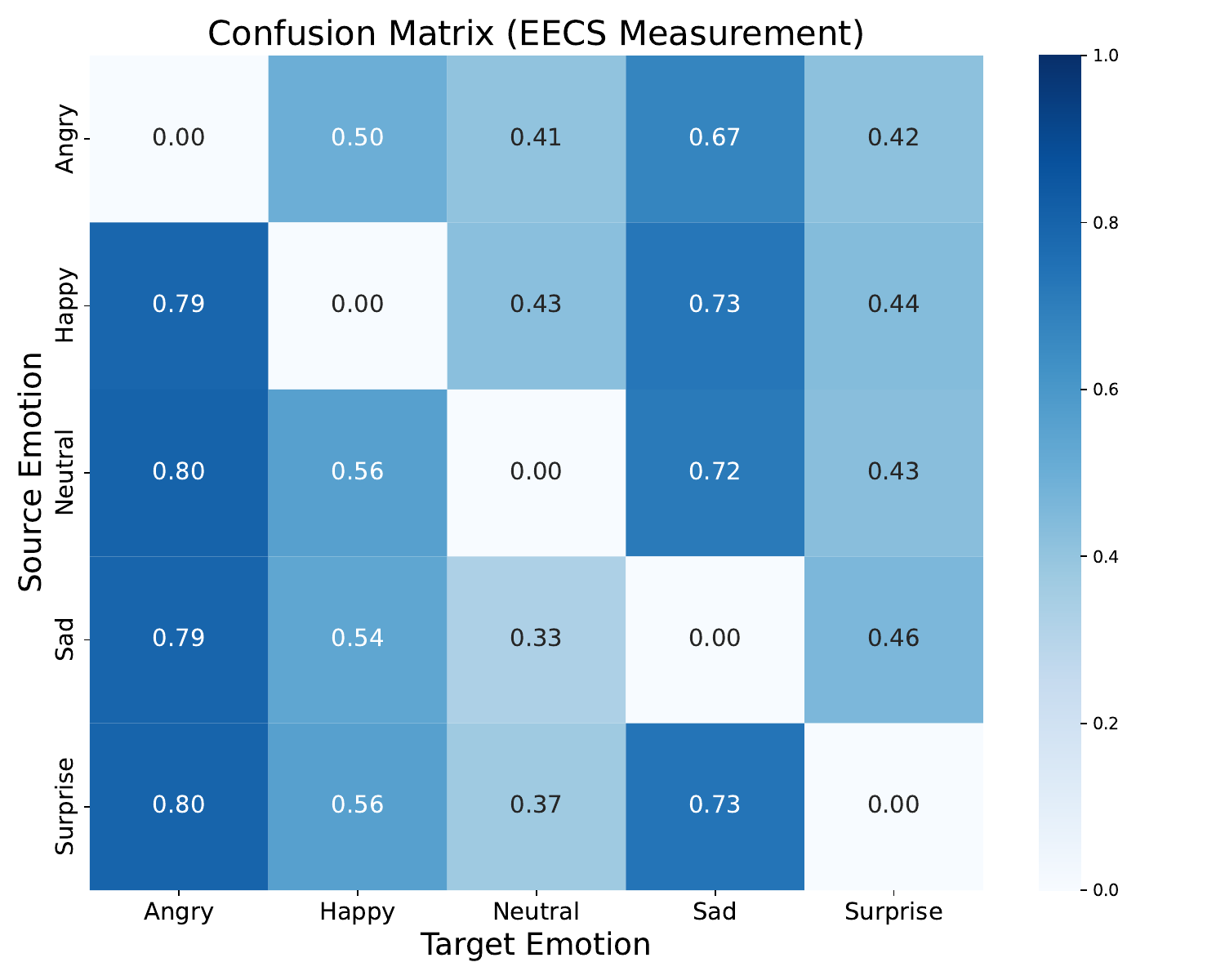}}
\subfloat[DurFlex-EVC]{\includegraphics[width=0.5\linewidth]{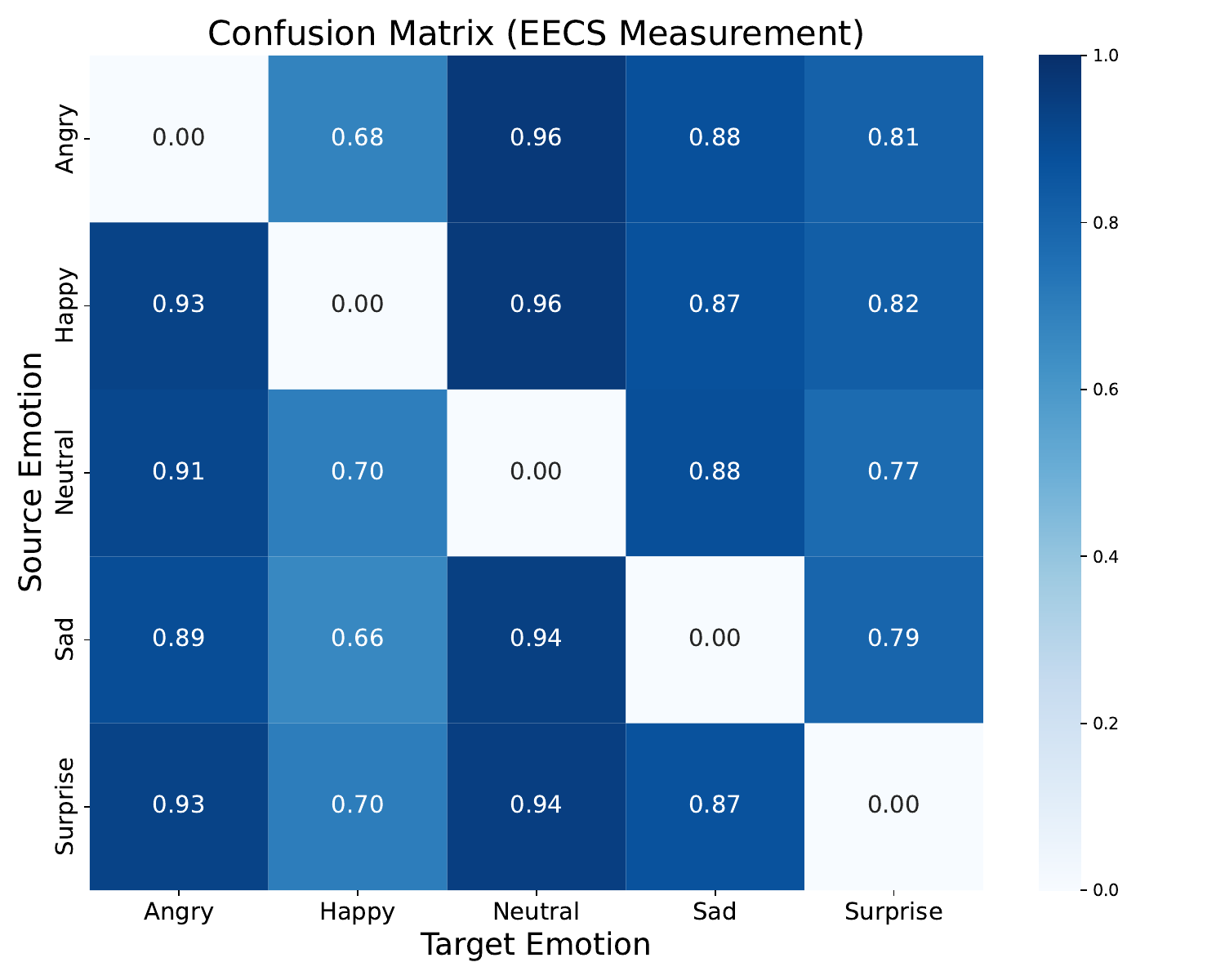}}
\caption{Comparison of the EECS scores of the comparison models for all combinations of emotion conversion.}
\label{fig:eecs_comparison}
\end{figure}

\subsubsection{Objective Metrics}
For our objective evaluation, we incorporate a range of metrics: predicted mean opinion score, phoneme error rate (PER), character error rate (CER), word error rate (WER), emotion classification accuracy (ECA), and speaker embedding cosine similarity (SECS). The predicted MOS was assessed using UTMOS \cite{saeki22c_interspeech}\footnote{\url{https://github.com/tarepan/SpeechMOS}}. 
For the PER calculation, we used a wav2vec2.0-based phoneme recognition model from Hugging Face \cite{xu22b_interspeech}. CER and WER were determined using Whisper\footnote{\url{https://github.com/openai/whisper}}\cite{pmlr-v202-radford23a}. In assessing SECS, we extracted speaker embeddings from both target and generated audio using Resemblyzer\footnote{\url{https://github.com/resemble-ai/Resemblyzer}}, subsequently computing their cosine similarity. This similarity measure ranges from -1 to 1, where higher values denote greater similarity. 
We evaluated the similarity for samples that shared the same speaker and emotion and then averaged these across all speakers. The objective evaluation of emotions in the generated speech was conducted using a pre-trained speech emotion recognition (SER) model. 

To measure SER accuracy, we employed emotion2vec \cite{ma2023emotion2vec}.
We used emotion2vec+ base\footnote{\url{https://github.com/ddlBoJack/emotion2vec}}, a pre-trained model that supports nine classes, and only used the five sentiment classes in the ESD dataset for evaluation.
We propose the emotion embedding cosine similarity (EECS) to evaluate the emotion of synthesized speech. 
The EECS is obtained by computing the cosine similarity of the emotion embedding between the synthesized audio and arbitrary reference audio with the target emotion. The emotion embedding was obtained using emotion2vec. Fig. \ref{fig:tsne} is a visualization of the features in emotion2vec, which shows that it encodes emotions independently of the speaker. We also evaluated the root mean square error (RMSE) for pitch and energy and calculated the difference of duration (DDUR) to assess prosody. The pitch was extracted using parselmouth\footnote{\url{https://parselmouth.readthedocs.io/en/stable/}} in Hz, and the energy was obtained as the L2-norm of the absolute value of the linear spectrogram. DDUR was obtained in the same way as in \cite{9778970}. We measured the real-time factor (RTF) of inference time for computational analysis.
Table \ref{tab:gt} presents the results of each pre-trained baseline model on the test set, comprising 1500 samples, for both ground-truth and vocoded samples.

\subsection{Comparison Models}
To benchmark the efficacy of our proposed method, we trained and compared it against several existing models.

\begin{itemize}
    \item StarGAN-EVC\footnote{\url{https://github.com/glam-imperial/EmotionalConversionStarGAN}}\cite{9054579}: 
    This adversarial network model specializes in speech emotion conversion. Its GAN-based architecture supports parallel generation, distinguishing it in this domain.
    \item Seq2seq-EVC\footnote{\url{https://github.com/KunZhou9646/seq2seq-EVC}}\cite{zhou21b_interspeech}: 
    Employing a seq2seq framework, this model adopts a two-stage strategy utilizing the TTS model. A notable feature of Seq2seq-EVC is its ability to jointly model duration and pitch. 
    \item Emovox\footnote{\url{https://github.com/KunZhou9646/Emovox}}\cite{9778970}: 
    Similar to Seq2seq-EVC, Emovox is based on a seq2seq structure. Its uniqueness lies in its focus on modulating emotional intensity. Emovox incorporates a ranking function to effectively model this intensity dimension. 
    \item Mixed-Emotions\footnote{\url{https://github.com/KunZhou9646/Mixed_Emotions}}\cite{10003644}: 
    Operating on a seq2seq framework similar to Emovox, this model is designed to express mixed emotions. It shares a controllable emotion intensity feature with Emovox.
    \item Textless-EVC\footnote{\url{https://github.com/facebookresearch/fairseq/tree/main/examples/emotion_conversion}}\cite{kreuk-etal-2022-textless}: 
    This model approaches speech synthesis by deconstructing the speech signal into discrete learned representations. These include speech content units, prosody features, speaker identity, and emotions. Each element is modified to align with the target emotion before being synthesized back into speech.
    \item DurFlex-EVC: 
    Our proposed model includes a unit aligner, style autoencoder, stochastic duration predictor, hierarchical stylize encoder, and a diffusion-based generator. This model stands out with its comprehensive and integrated approach to emotional speech synthesis.
\end{itemize}
All comparison models were trained using the official implementation. We used the same vocoder to generate the waveforms except for Textless-EVC, which generates the waveform directly, and adjusted the hyperparameters to match the vocoder settings. 

\begin{table}[!t]
\centering
\caption{Results of subjective evaluations each comparison model}
\label{tab:subjective}
\begin{tabular}{l|ccc}
\toprule
    Model & nMOS & sMOS & eMOC  \\
    \midrule
    GT              & 3.72 ($\pm$0.03) & 3.95 ($\pm$0.06) & 82.98 \\
    GT (vocoded)    & 3.70 ($\pm$0.05) & 3.58 ($\pm$0.11) & 82.98 \\
    \midrule
    StarGAN-EVC     & 3.59 ($\pm$0.06) & 3.36 ($\pm$0.12) & 37.84 \\ 
    Seq2seq-EVC     & 3.43 ($\pm$0.07) & 3.09 ($\pm$0.13) & 48.65 \\ 
    Emovox          & 3.50 ($\pm$0.06) & 3.10 ($\pm$0.13) & 51.35 \\ 
    Mixed Emotion   & 3.50 ($\pm$0.07) & 3.27 ($\pm$0.12) & 62.16 \\ 
    Textless-EVC    & 3.61 ($\pm$0.05) & 3.39 ($\pm$0.11) & 56.76 \\ 
    DurFlex-EVC     & 3.70 ($\pm$0.05) & 3.63 ($\pm$0.10) & 72.97 \\ 
\bottomrule
\end{tabular}
\end{table}

\begin{table*}[!t]
\centering
\caption{Results of objective evaluations for each comparison model} 
\label{tab:objective}
\resizebox{0.8\linewidth}{!}{
\begin{tabular}{l|cccccccc}
\toprule
        Model & UTMOS & PER & CER & WER & ECA &  EECS & SECS & RTF \\
        \midrule
        StarGAN-EVC     & 1.47 & 70.83 & 44.49 & 67.71 & 39.5  & 0.34 & 0.61 & 0.0051\\ 
        Seq2seq-EVC     & 1.54 & 37.29 & 21.68 & 36.87 & 40.0  & 0.39 & 0.63 & 0.2325\\ 
        Emovox          & 2.05 & 29.25 & 17.18 & 31.37 & 49.33 & 0.48 & 0.68 & 0.2109\\ 
        Mixed Emotion   & 2.02 & 29.86 & 18.21 & 33.09 & 57.75 & 0.55 & 0.67 & 0.2289\\ 
        Textless-EVC    & 2.37 & 22.88 & 12.49 & 23.98 & 56.18 & 0.58 & 0.68 & 0.4340\\ 
        DurFlex-EVC (4 time steps)     & 3.21 & 18.89 & 8.61 & 21.36 & 89.52 & 0.85 & 0.71 & 0.1334\\ 
        DurFlex-EVC (100 time steps)     & 3.39 & 17.31 & 8.26  & 20.75 & 88.64 & 0.85 & 0.75 & 1.9200 \\ 
\bottomrule
\end{tabular}    
}
\end{table*}
\section{Result}
This section contains the results and discussion of the extensive experiments. We compared our proposed model with previous EVC models to evaluate its quality. We then conducted experiments to demonstrate the effectiveness of the components of the model. Furthermore, we conducted extended experiments on unseen speaker scenarios. 

\subsection{Comparison of evaluation results for baseline models}
To evaluate the performance of the proposed model, we conducted both objective and subjective evaluations to compare it with the baseline models. Table \ref{tab:subjective} shows the results of subjective evaluation. 
Table \ref{tab:objective} shows the objective evaluation results. As the nMOS results show, the naturalness of the speech generated by our method outperforms that of other models. In terms of speaker similarity, our approach scored the highest on both sMOS and SECS. Fig. \ref{fig:secs_comparison} shows the resulting SECS for all combinations of emotional conversions. This demonstrates that the proposed model is more robust in terms of speaker similarity than the comparison models for all combinations of transformations. Furthermore, the results of eMOC, ECA and EECS demonstrate that our method performs better in terms of perceptual quality as well as objective metrics. Fig. \ref{fig:eecs_comparison} shows the resulting EECS for all combinations of emotion transformations. This indicates that the proposed model synthesized speech with a higher emotional similarity than the comparison models for all combinations of transformations. The ASR evaluation also shows that our method achieves lower values in PER, CER, and WER compared to other models. This emphasizes the ability of our method to synthesize precisely pronounced speech.

We evaluated the computational efficiency of each model by comparing their RTF. Autoregressive models, including Seq2seq-EVC, Emovox, Mixed Emotion, and Textless-EVC, generally demonstrated slower inference speeds compared to parallel generation models such as StarGAN-EVC. DurFlex-EVC, as a diffusion-based model, exhibited varying RTFs depending on the number of sampling time steps. With just four time steps, DurFlex-EVC achieved an RTF of 0.1334, significantly faster than all compared autoregressive models. At this setting, DurFlex-EVC maintained competitive performance and exceeded the speed and performance metrics of other autoregressive models. However, when using 100 time steps, DurFlex-EVC achieved the highest performance, though at a slower RTF of 1.9200 due to the increased computational demand of the diffusion process.

For a quantitative assessment of prosody, we also compared pitch and energy duration. Table \ref{prosody} shows the evaluation results for each emotion. We found that the proposed model scored better than the other models in all prosody evaluations.

\subsection{Experiments for Analyzing Model Architectures}
We conducted an analysis of the model design and additional experiments. Table \ref{ablation} includes the results of the ablation study and additional experiments. 

\subsubsection{Effectiveness of Style Autoencoder}
\begin{figure}[!t]
    \centering
    
    \includegraphics[width=1.0\linewidth]{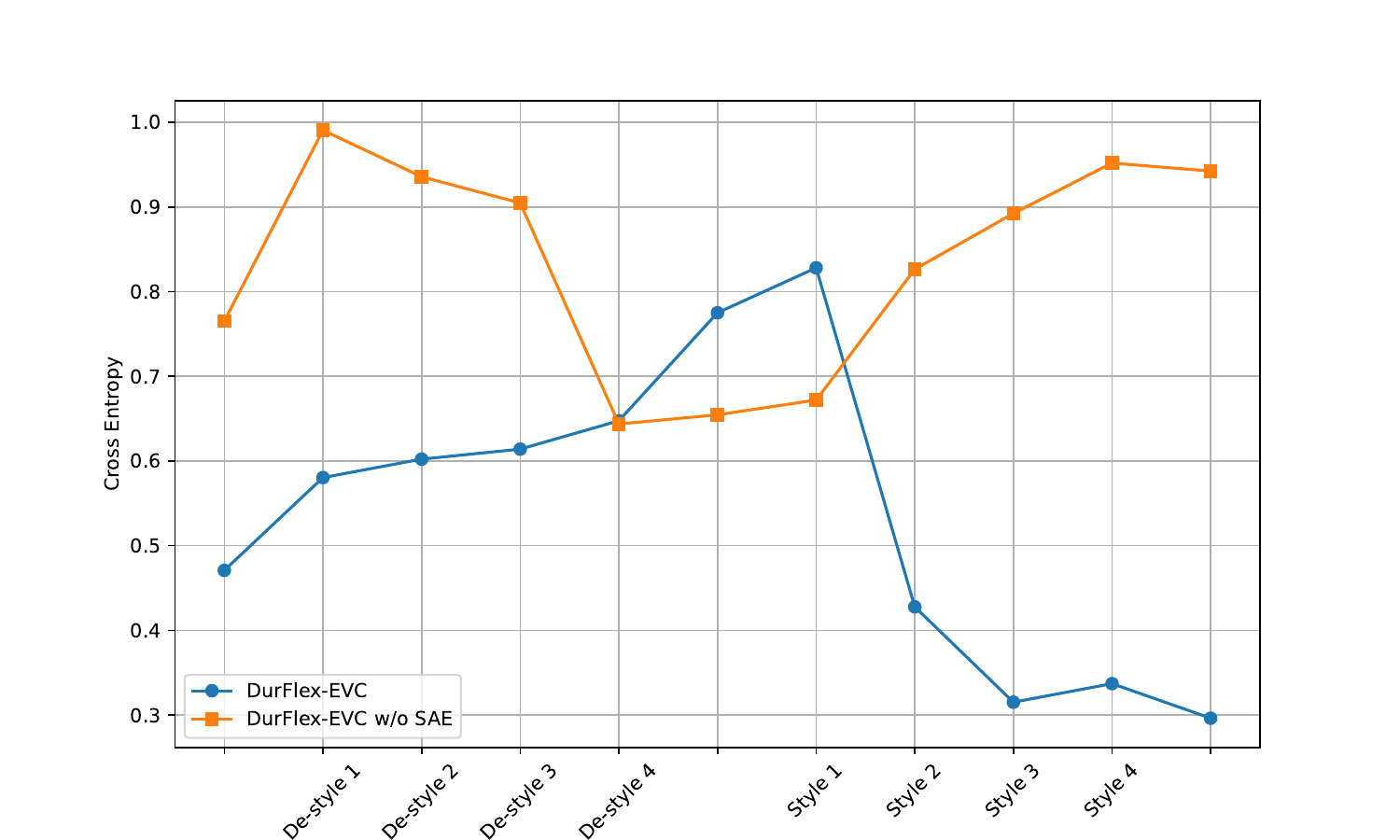}
    \caption{Cross-entropy change across layers for emotion classification. Models with the SAE show increasing cross-entropy through de-stylize transformers and decreasing cross-entropy through stylize transformers. Models without the SAE show consistently high cross-entropy.}
    \label{fig:cross_entropy}
\end{figure}

We conducted an experiment to verify the effectiveness of the style autoencoder (SAE). \textit{w/o SAE} is a model that stacks a standard feed forward transformer block with a full layer of SAE, without perturbation and conditioning for emotion style. In Table \ref{ablation}, \textit{w/o SAE} shows the results of the ablation study on it. The overall performance degradation observed in models without SAE indicates the importance of disentangling content and style from input features. Fig. \ref{fig:cross_entropy} shows the change in cross entropy for styles as input features pass through SAE. To obtain cross entropy, we trained a classifier for emotion style on features extracted from all layers of SAE. The black line is the result for the model with SAE and the orange line is the result for the model without SAE. The \textit{w/o SAE} model shows high cross entropy across the layers, while the model with SAE shows increasing cross entropy as it passes through the de-stylize transformer layers and decreasing cross entropy as it passes through the stylize transformer layers. This means that the style is disentangled and conditioned from the feature by SAE.

We also experimented with adversarial training strategies to remove source styles from input features. This replaces the de-stylize transformer in the style autoencoder with a standard transformer, and adds a gradient reversal layer and classifier to prevent its output from learning the style. This adversarial training strategy is often used to disentangle specific styles \cite{wang21h_interspeech, 9747987, 10106091}. 

The results for this are shown in Table \ref{ablation} as \textit{w/ adv}, and show a performance degradation on all metrics except EECS. 
\begin{table}[!t]
\centering

\caption{Comparison of prosody features (pitch, energy, duration).}
\resizebox{1.0\linewidth}{!}{
\begin{tabular}{lccccc|c}
\toprule
\multicolumn{7}{c}{Pitch (RMSE)} \\
\midrule
Model & Neutral & Angry & Happy & Sad & Surprise & Avg.\\
\midrule
StarGAN-EVC         & 54.96 & 46.5 & 59.14 & 53.51 & 68.69 & 56.56 \\ 
Seq2Seq-EVC         & 62.03 & 56.6 & 70.78 & 55.21 & 65.09 & 61.94 \\ 
Emovox              & 53.35 & 49.35 & 57.18 & 51.42 & 52.96 & 52.85 \\ 
Mixed-Emotion       & 55.0 & 49.24 & 56.92 & 51.63 & 55.01 & 53.56 \\ 
Textless-EVC        & 57.15 & 47.85 & 51.09 & 54.86 & 48.72 & 51.93 \\ 
DurFlex-EVC         & 49.43 & 45.79 & 50.53 & 47.46 & 54.39 & 49.52 \\ 
\midrule
\multicolumn{7}{c}{Energy (RMSE)} \\
\midrule
Model & Neutral & Angry & Happy & Sad & Surprise & Avg.\\
\midrule
StarGAN-EVC         & 21.00 & 19.80 & 20.17 & 20.92 & 19.54 & 20.28 \\ 
Seq2Seq-EVC         & 25.46 & 24.54 & 24.30 & 24.67 & 24.32 & 24.66 \\ 
Emovox              & 25.22 & 24.64 & 24.02 & 24.73 & 24.01 & 24.53 \\ 
Mixed-Emotion       & 25.46 & 24.20 & 23.74 & 24.40 & 23.56 & 24.27 \\ 
Textless-EVC        & 13.44 & 13.52 & 14.32 & 13.96 & 13.88 & 13.82 \\ 
DurFlex-EVC         & 12.25 & 12.31 & 12.64 & 13.16 & 12.63 & 12.60 \\
\midrule
\multicolumn{7}{c}{Duration (DDUR)} \\
\midrule
Model & Neutral & Angry & Happy & Sad & Surprise & Avg.\\
\midrule
StarGAN-EVC         & 0.31 & 0.34 & 0.26 & 0.26 & 0.28 & 0.29\\
Seq2Seq-EVC         & 0.21 & 0.22 & 0.27 & 0.22 & 0.23 & 0.23 \\
Emovox              & 0.22 & 0.23 & 0.23 & 0.21 & 0.24 & 0.23 \\
Mixed-Emotions      & 0.22 & 0.22 & 0.29 & 0.25 & 0.30 & 0.26 \\
Textless-EVC        & 0.30 & 0.36 & 0.34 & 0.28 & 0.33 & 0.32 \\
DurFlex-EVC         & 0.20 & 0.21 & 0.24 & 0.23 & 0.23 & 0.22 \\
\bottomrule
\label{prosody}
\end{tabular}
}
\end{table}

\subsubsection{Effectiveness of Unit Aligner}
The unit aligner (UA) was introduced to model stylized contexts. Table \ref{ablation} \textit{w/o UA} shows the results of the ablation experiment for this. The results showed that \textit{w/o UA} performed better in terms of voice quality and pronunciation but not in terms of emotion conversion. This indicates that there remains a lot of information about the source style in the features and that the UA functions as a bottleneck and has a significant impact on style control. We measured the Jensen-Shannon divergence (JSD) to validate that the predicted units in UA reflect the target emotion. Table \ref{table:jsd_units} shows the JSD results of the units for each emotion in Textless-EVC and DurFlex-EVC. DurFlex-EVC achieved a better JSD than Textless-EVC for all emotions, indicating that it provides an appropriate representation of the context. 

\subsubsection{Effectiveness of Hierarchical Stylize Encoder}
The output of the unit aligner is a representation in units, which is expanded to frame-level by a duration predictor and length regulator. For the frame-level stylization in this representation, we introduce a hierarchical stylize encoder (HSE). The results of the ablation study without the HSE are shown in Table \ref{ablation} under the entry labeled \textit{w/o HSE}. The evaluation results showed an overall performance degradation with \textit{w/o HSE}. This is inferred to be due to HSE reducing the burden on the Mel-spectrogram generator, resulting in improved generation quality. 

\begin{figure}[!t]
    \centering
    
    \subfloat[Angry]{\includegraphics[width=0.45\linewidth]{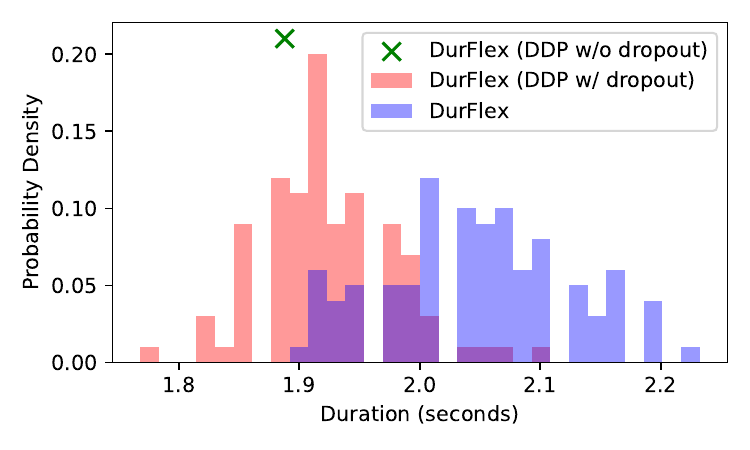}}
    \subfloat[Happy]{\includegraphics[width=0.45\linewidth]{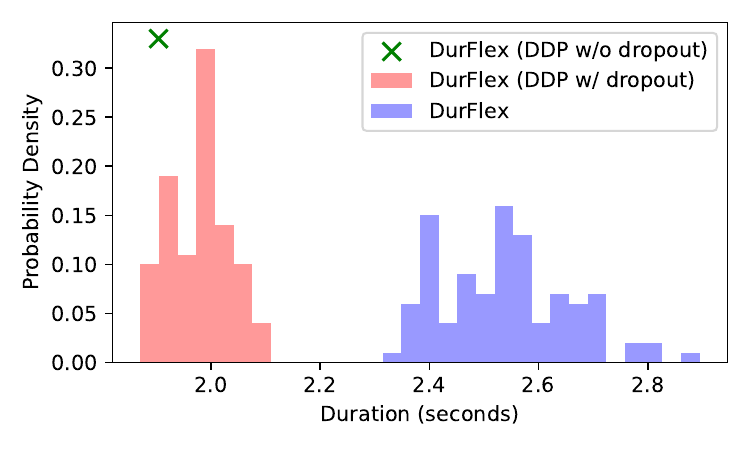}}

    \subfloat[Sad]{\includegraphics[width=0.45\linewidth]{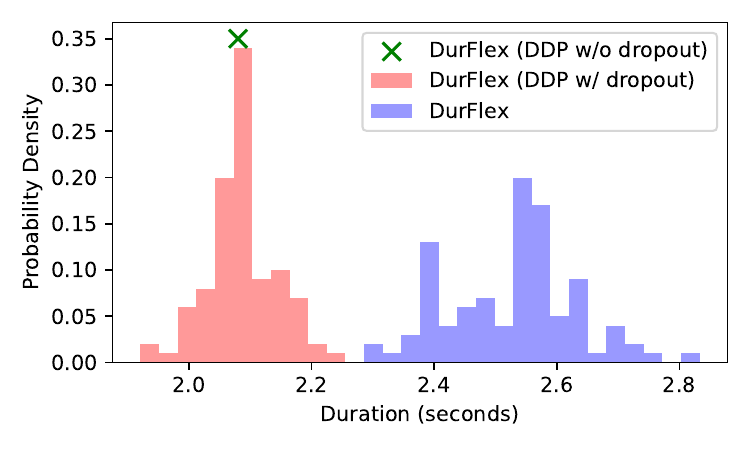}}
    \subfloat[Surprise]{\includegraphics[width=0.45\linewidth]{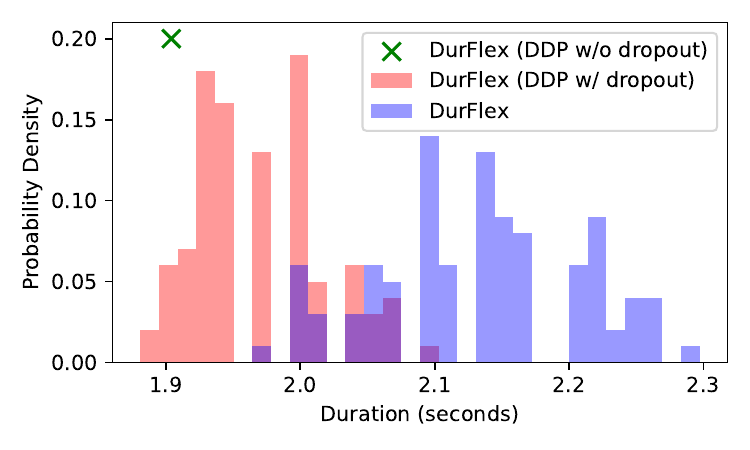}}

    \caption{Histogram of duration by emotion based on the structure of the duration predictor.}
\label{dur_hist}
\end{figure}

\begin{table}[!t]
\centering
\caption{Results of ablation studies and additional experiments}
\label{ablation}
\resizebox{1.0\linewidth}{!}{
\begin{tabular}{lccccccc}
\toprule
        Model                       & UTMOS & PER & CER & WER & ECA & EECS & SECS \\
        \midrule
        DurFlex-EVC                 & 3.39 & 17.31 & 8.26 & 20.75 & 88.64 & 0.85 & 0.75  \\
        \midrule
        w/o SAE                     & 3.34 & 18.28 & 9.37 & 22.64 & 87.64 & 0.83 & 0.72 \\ 
        w/o UA                      & 3.55 & 12.31 & 3.55 & 13.04 & 24.39 & 0.31 & 0.66 \\ 
        w/o HSE  & 3.27 & 20.00 & 9.32 & 22.65 & 68.95 & 0.65 & 0.69 \\ 
        w/ DDP                      & 3.39 & 17.60 & 8.59 & 21.56 & 87.52 & 0.83 & 0.73  \\ 
        w/ FFT                      & 3.03 & 17.32 & 7.42 & 19.41 & 54.68 & 0.57 & 0.73  \\ 
        w/ adv   & 3.38 & 18.47 & 8.33 & 21.18 & 87.53 & 0.84 & 0.71 \\
        w/ unit2mel              & 3.30 & 20.18 & 9.32 & 22.65 & 89.11 & 0.85 & 0.69 \\ 
        w/ unit2wav    & 1.26 & 18.86 & 7.69 & 18.56 & 20.23 & 0.29 & 0.51 \\ 
\bottomrule
\end{tabular}    
}
\end{table}

\begin{table}[!t]
\centering
\caption{JSD of unit distributions by emotion}
\begin{tabular}{l|ccccc}
\toprule
& Neutral & Angry & Happy & Sad & Surprise \\
\midrule
Textless-EVC & 0.21 & 0.20 & 0.22 & 0.22 & 0.21 \\ 
DurFlex-EVC & 0.11 & 0.11 & 0.12 & 0.12 & 0.11 \\ 
\bottomrule
\end{tabular}
\label{table:jsd_units}
\end{table}

\subsubsection{Comparison for Duration Predictors}
We introduced a stochastic duration predictor (SDP) for duration modeling to represent the diversity of emotions. To evaluate, we compared our method with widely used deterministic duration predictors (DDP). The DDP follows the structure described in FastSpeech \cite{NEURIPS2019_f63f65b5}, which includes two convolutional layers, ReLU activation, layer normalization, and dropout. Tables \ref{ablation}, \textit{w/ DDP} shows the evaluation results of the model using DDP. The evaluation results show that the model using SDP is better than the model using DDP. We compared the JSD of unit duration for each emotion for each duration predictor. Table \ref{table:jsd_duration} shows the JSD results for unit duration for models with stochastic duration predictors, \textit{w/ SDP}, and deterministic duration predictors, \textit{w/ DDP}.

In our experiments, we found that DDP does not always output the same length when generating the same sentence with the same conditions. We discovered that it was caused by a dropout within the DDP. We experimented with repeatedly generating the same speech, and Fig. \ref{dur_hist} shows a histogram of the duration of the speech for each emotion. Red bars represent DDP, black bars represent SDP, and 'x' represents the results for DDP with dropout removed. Table \ref{table:jsd_duration}, \textit{w/ DDP (w/o dropout)}, shows the results for a fully deterministic duration predictor without dropout, which gave almost similar results to the version with dropout. SDP showed better JSD than \textit{w/ DDP} and \textit{w/ DDP (w/o dropout)}, indicating that it is suitable for modeling emotional duration distributions. 

\begin{figure}[!t] 
    \centering
    \subfloat[DurFlex-EVC w/ FFT]{\includegraphics[width=1.0\linewidth]{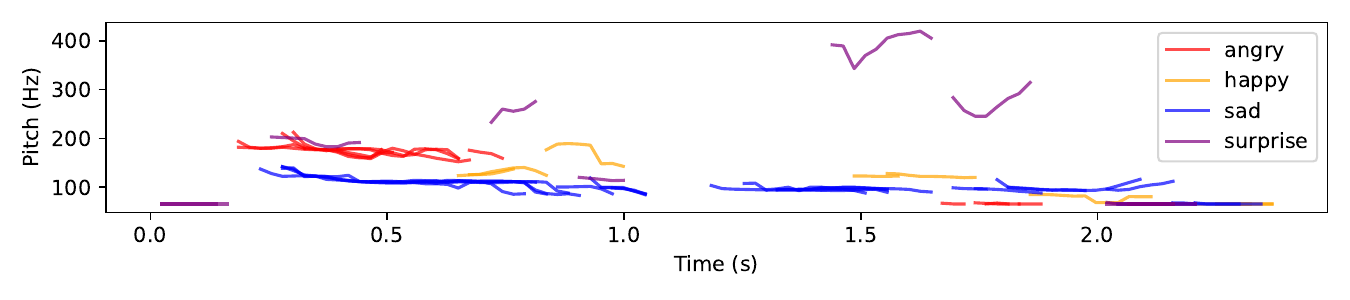}
    \label{pitch_fft}}
    
    \subfloat[DurFlex-EVC w/ diffusion]{\includegraphics[width=1.0\linewidth]{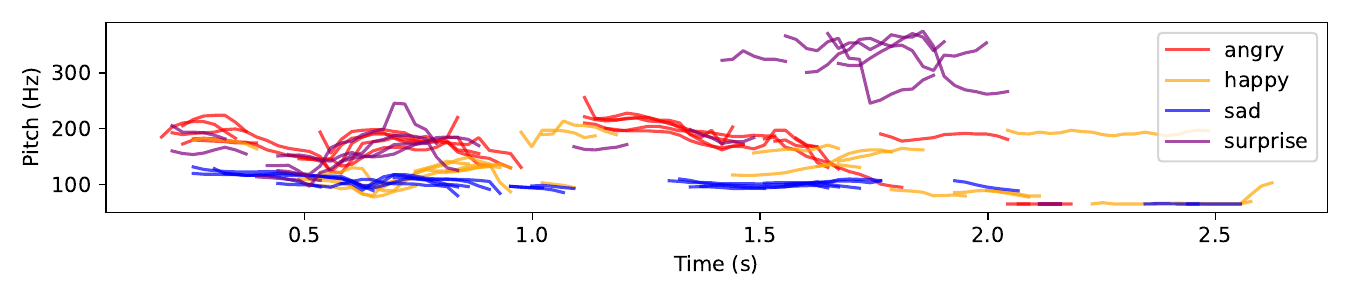}
    \label{pitch_diff}}
    \caption{Pitch track of the same speech converted multiple times for each emotion.}
\label{pitch_track}
\end{figure}

\begin{table}[!t]
\centering
\caption{JSD for unit durations by emotion}
\resizebox{1.0\linewidth}{!}{%
\begin{tabular}{l|ccccc}
\toprule
Model & Neutral & Angry & Happy & Sad & Surprise \\
\midrule
w/ DDP (w/o dropout)    & 0.037 & 0.021 & 0.021 & 0.064 & 0.021 \\
w/ DDP                  & 0.037 & 0.021 & 0.022 & 0.063 & 0.022 \\
w/ SDP                  & 0.027 & 0.015 & 0.017 & 0.050 & 0.021 \\
\bottomrule
\end{tabular}
}
\label{table:jsd_duration}
\end{table}

\begin{table}[!t]
\centering
\caption{JSD for pitch by emotion}
\begin{tabular}{l|ccccc}
\toprule
Model           & Neutral & Angry & Happy & Sad & Surprise \\
\midrule
w/ FFT          & 0.21 & 0.13 & 0.15 & 0.17 & 0.24 \\
w/ diffusion    & 0.19 & 0.11 & 0.16 & 0.15 & 0.25 \\
\bottomrule
\end{tabular}
\label{table:jsd_pitch}
\end{table}
\subsubsection{Comparison for Mel-spectrogram Generator Structures}
Diffusion-based generators have been shown to produce high-quality and diverse results in a wide range of domains. We introduced a diffusion-based structure to generate more expressive speech for each emotion. We conducted experiments to compare our decoder with a feed-forward transformer (FFT) based structure, which is widely used in conventional speech synthesis studies for parallel generation. Table \ref{ablation} \textit{w/ FFT} shows the objective evaluation results of the model using the FFT-based decoder. The \textit{w/ FFT} scored better in CER and WER for pronunciation, but lower on UTMOS for quality, and worse on ECA and EECS for emotion expression.

To verify the expressiveness of each emotion, we computed the JSD over pitch for each emotion. In Table \ref{table:jsd_pitch}, \textit{w/ FFT} is the model using an FFT decoder and \textit{w/ diffusion} is the model using a diffusion-based structure. \textit{w/ diffusion} shows better JSD on neutral, angry, sad, while \textit{w/ FFT} shows better JSD in happy and surprise. Although \textit{w/ FFT} outscores \textit{w/ diffusion} in Happy and Surprise, it does not mean that FFT is more expressive. Fig. \ref{pitch_track} shows the pitch of the speech generated by multiple iterations of the same sentence. Models using FFT produce consistent pitch tracks that remain stable over repeated generations, while models using diffusion produce more dynamic results. This suggests that the diffusion-based structure captures higher expressiveness, resulting in a higher JSD compared to the FFT-based structure, which tends to average over dynamic emotions like happy and surprise.

\begin{figure}[!t] 
    \centering 
    \includegraphics[width=0.8\linewidth]{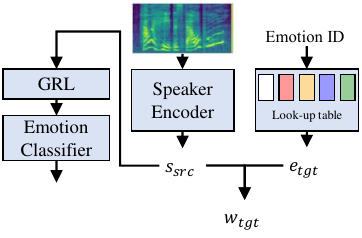}
    \caption{Style embedding modeling for unseen speaker setting.}
\label{style_embedding}
\end{figure}

\begin{table}[!t]
\centering
\caption{Comparison of results based on input features} 
\resizebox{1.0\linewidth}{!}{
\begin{tabular}{lccccccc}
\toprule
Model                       & UTMOS & PER & CER & WER & ECA & EECS & SECS \\
\midrule
w/ Mel-spec.                & 3.29 & 25.28 & 15.70 & 31.31 & 88.10 & 0.84 & 0.72 \\ 
w/ linear-spec.             & 3.29 & 25.98 & 16.29 & 31.81 & 89.88 & 0.85 & 0.68 \\ 
w/ wav2vec 2.0              & 3.34 & 30.55 & 21.46 & 38.44 & 91.03 & 0.86 & 0.66 \\ 
w/ wavLM                    & 3.36 & 23.44 & 12.07 & 26.22 & 92.59 & 0.87 & 0.67  \\ 
w/ HuBERT                   & 3.39 & 17.31 & 8.26  & 20.75 & 88.64 & 0.85 & 0.75 \\
\bottomrule
\end{tabular}    
}
\label{input_representation} 
\end{table}

\subsubsection{Comparison for Input Features}
The proposed model takes HuBERT features as input and outputs a Mel-spectrogram. The reason for outputting Mel-spectrogram is for compatibility with pre-trained vocoders. To explain the reasoning behind our choice of HuBERT features, we present a comparison of the input features. We compared Mel-spectrogram and linear spectrogram as features using conventional signal processing, and wav2vec 2.0 \cite{Baevski2020vq-wav2vec}, wavLM \cite{9814838}, and HuBERT \cite{9585401} as SSL features that have recently been used as linguistic representations \cite{NEURIPS2022_69c754f5, lee23i_interspeech}.

Although the input feature dimensions differ in each experiment, they are aligned to the same hidden dimension through a linear layer. 

Table \ref{input_representation} shows the results of the experiments by input feature. The results show that models using Mel-spectrogram and linear-spectrogram have lower UTMOS than SSL feature models. Models using wavLM performed better overall than those using wav2vec 2.0. The model using HuBERT outperformed the others in all metrics except ECA and EECS. In particular, the model using HuBERT shows a significant improvement in pronunciation, which is interpreted as advantageous for linguistic learning over other features because the target units used for training are obtained from the clustering of HuBERT features. To compare the linguistic modeling ability of each input feature, we calculated the BLEU score \cite{10.3115/1073083.1073135} and the unit error rate (UER) of the predicted units. Table \ref{bleu} shows the BLEU score and UER of the predicted units for each feature. We found that the prediction of units was correlated with the pronunciation accuracy of the synthesized speech, and this was due to the fact that HuBERT predicted units more accurately than the other features.

We also experimented with a model that takes unit input and generates speech. In Table \ref{ablation}, \textit{w/ unit2mel} is a model that receives unit input and generates Mel-spectrogram. This model has the same structure, with the addition of a unit embedding layer to accept units as input. Furthermore, we experimented with models that generate waveforms directly from unit. \textit{w/ unit2wav} is a modification of HiFi-GAN \cite{NEURIPS2020_c5d73680} to take unit as input, which is equivalent to not including f0 in Textless-EVC generator. 

The \textit{w/ unit2mel} shows a worse overall performance, except for ECA and EECS. The \textit{w/ unit2wav} shows significantly worse metrics across the board, except for CER and WER. We interpret this to mean that the speech unit has enough information about pronunciation but not enough other speech information to generate a waveform. To overcome this, Textless-EVC uses additional information such as pitch and timbre to generate the waveform. 

\begin{figure}[!t] 
    \centering 
    \includegraphics[width=1.0\linewidth]{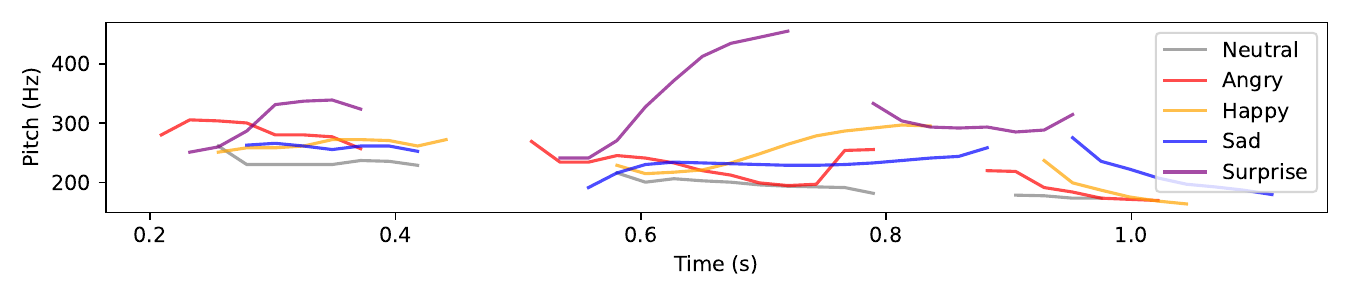}
    \caption{Pitch track of unseen speaker speech converted to each emotion.}
\label{pitch_zero}
\end{figure}

\begin{table}[!t]
\centering
\caption{Comparison of BLEU score and UER based on input features}
\begin{tabular}{lcc}
\toprule
Model                       & BLEU & UER \\
\midrule
w/ Mel-spec.                & 15.23 & 62.62 \\ 
w/ linear-spec.             & 15.66 & 60.93 \\ 
w/ wav2vec 2.0              & 11.96 & 62.67 \\ 
w/ wavLM                    & 25.51 & 46.43 \\ 
w/ HuBERT                   & 38.59 & 38.04 \\
\bottomrule
\end{tabular}    
\label{bleu}
\end{table}
\subsection{Unseen Speaker Emotion Conversion} 
We extended our experiments to apply our proposed model to an unseen speaker scenario. To make it possible, we modified the model structure to allow speaker information to encode speaker embedding from reference audio instead of speaker IDs. We adopt the style encoder structure from Meta-StyleSpeech \cite{pmlr-v139-min21b} as the speaker encoder. We added a gradient reversal layer (GRL) \cite{pmlr-v37-ganin15} and a linear layer to prevent the speaker encoder from learning information about emotion. The linear layer performs the emotion classification task, where the losses are reversed by the GRL to prevent the speaker encoder from learning about emotion. Fig. \ref{style_embedding} shows the speaker encoder and emotion embedding designed in this way to model style embedding. We set the weight for adversarial losses due to GRL to 0.001. We trained our model on the ESD dataset. For the unseen speaker test, we composed the test set by randomly selecting five sentences for each speaker from the VCTK dataset that were not used for training. We set the emotion of the test set to neutral and converted all other emotions.

\begin{figure}[!t] 
    \centering 
    
    \includegraphics[width=0.7\linewidth]{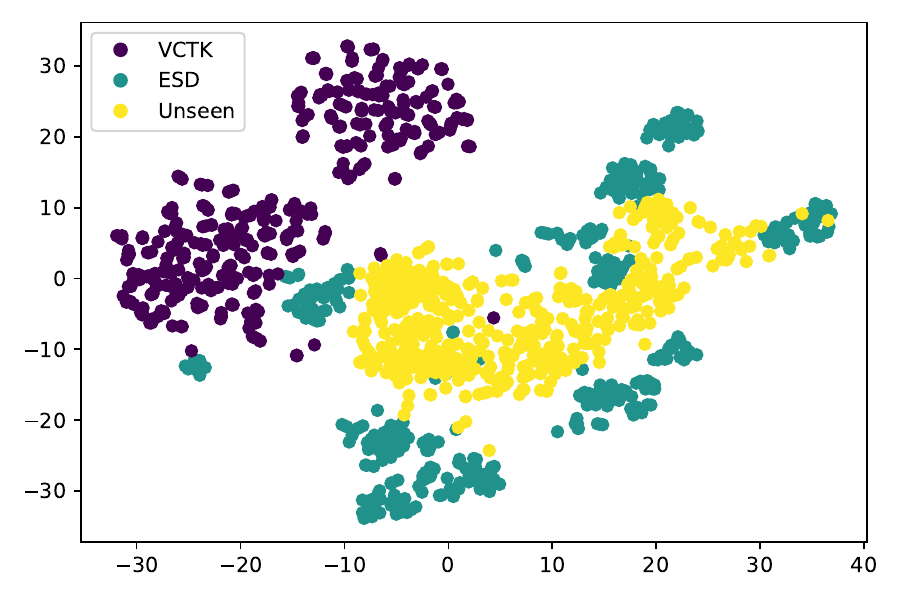}
    \caption{Visualization of t-SNEs of speaker embedding for the ESD dataset (green), the VCTK dataset (purple) and unseen speaker test results (yellow).}
    
\label{tsne_unseen}
\end{figure}

Table \ref{zeroshot} shows the evaluation results for seen and unseen speakers for the modified model. The modified model scored better UTMOS, PER, CER, and WER than the original version, while performing weaker on ECA, EECS, and SECS. We infer that the synthesis quality and pronunciation are better due to the additional information encoded from the reference audio that helps with speech synthesis. However, the lack of style disentanglement leads to decreased performance on emotion and speaker-related metrics. The results of the unseen speaker show that the modified structure allows for the emotion conversion of a new speaker without losing quality. Fig. \ref{pitch_zero} shows the pitch tracks of the transformed samples of the unseen speaker for each emotion, showing distinct differences for each emotion. We found that the speaker similarity for unseen speakers was lower compared to seen speakers. Table \ref{table:unseen_secs} shows the speaker similarity for each emotion. We observe a decrease in the speaker similarity for all emotions. Fig. \ref{tsne_unseen} shows a t-SNE visualization of speaker embedding for the ESD dataset (green), the VCTK dataset (purple), and the results of the unseen speaker experiment (yellow). The model synthesized speech that was closer to the speaker in the ESD, which was the training set. 

\section{Discussion}
Our model is influenced by the framework described in \cite{8341805}, which focuses on modeling emotional pronunciation. However, our approach differs substantially; whereas their model directly addresses pronunciation, our focus is on converting SSL features at the unit level. This perspective aligns more closely with the emotional translation mechanisms inherent in Textless-EVC \cite{kreuk-etal-2022-textless}. The primary distinction in our approach lies in the utilization of the cross-attention output as the input for our model, rather than relying on the predicted units.

We designed our model to generate a Mel-spectrogram. This was more efficient than generating the waveform directly, and allowed us to do more experimentation. For example, Textless-EVC, which generates waveforms directly, spent two weeks training, while our model required three days.

\subsection{Limitations}
In experimental results, the diffusion-based model demonstrated improved results. However, the limitations of diffusion-based structures include their extensive computational demand and time-consuming nature. We anticipate that this challenge will be alleviated by the advent of the recent fast sampling method\cite{xiao2022tackling}. Some applications have been developed in speech research \cite{10517426}. The scope of our experiments also encompassed speaker generalization. Although we successfully observed emotional transformations in the voices of unseen speakers, a discernible lack of speaker similarity was apparent, indicating the need for further refinement. Empirically, our observations suggest that models tested on unseen data often reflect the voice distribution of the data used for their pre-training. Expanding the dataset is expected to increase the representational capacity of the generator, and the performance of zero-shot emotion conversion is also expected to improve. In the configuration of our style autoencoder, we utilized MixLN for de-stylizing and CLN for stylizing. It was observed that the perturbations introduced by MixLN resulted in a compromise that affected the equilibrium between expressiveness and pronunciation accuracy within our model. In addition, the task of effectively separating style from content remains a significant challenge \cite{Lee_Yoon_Noh_Kim_Lee_2021, NEURIPS2021_0266e33d}, requiring ongoing research and development to address this issue.

Although the diffusion model produces high-quality speech, it has limitations due to emotion datasets typically having a 16k sampling rate. Audio super-resolution models, such as \cite{10381805}, are expected to solve this problem. Although we only experimented with English, our model has the potential to be extended by considering a wide range of languages and combining it with speech-to-speech translations \cite{10447331}. The expression of emotions differs between people, languages, and cultures, and research is needed to reflect these differences. 

\begin{table}[!t]
\centering
\caption{Evaluation results for seen and unseen speakers} 
\label{table:unseen_secs}
\resizebox{1.0\linewidth}{!}{
\begin{tabular}{lccccccc}
\toprule
Model                       & UTMOS & PER & CER & WER & ECA & EECS & SECS \\
\midrule
GT (Seen)              & 3.60 & 11.64 & 3.06 & 12.09 & 89.46 & 0.76 & 0.81 \\
GT (Unseen)            & 4.03 & 10.07 & 0.67 & 1.39  & -                        & -                       & 0.84 \\
\midrule
Seen                   & 3.44 & 16.72 & 7.75 & 20.16 & 82.92 & 0.75 & 0.66 \\
Unseen                 & 3.53 & 18.83 & 7.35 & 12.18 & 75.85 & 0.72 & 0.60 \\
\bottomrule
\end{tabular}    
}
\label{zeroshot}
\end{table}

\begin{table}[!t]
\centering

\caption{SECS for seen \& unseen speaker emotional conversions}
\begin{tabular}{l|ccccc}
\toprule
Dataset & Angry & Happy & Sad & Surprise & Avg. \\
\midrule
Seen & 0.67 & 0.65 & 0.66 & 0.65 &  0.66 \\
Unseen & 0.63 & 0.58 & 0.60 & 0.59 &  0.60\\
\bottomrule
\end{tabular}
\end{table}
\subsection{Future Works}
Fundamentally, the proposed model depends on the performance of the SSL model because it utilizes discrete units. In addition to semantic units such as HuBERT, we also plan to investigate structures that exploit neural audio codecs such as \cite{dfossez2023high, 9625818}. Recent research in the field of speech synthesis has increasingly focused on controlling emotion intensity. Some works, such as \cite{9003829} and \cite{9383524}, have adopted the method of modeling emotion intensity using relative attribute ranking functions. Alternatively, some studies, such as \cite{9053732}, have explored the modeling of intensity through interpolation of embeddings. Various approaches, such as \cite{9747098}, have tried to control the intensity of emotions. 
Although these studies have shown that modeling emotion intensity is possible, precise control of intensity remains a challenge.
In our future work, we will explore methods to control emotional intensity. We will also investigate emotional voice conversion for cross-language. Furthermore, it is expected to be applied to singing voice synthesis tasks \cite{10447981} to express emotions. 

\section{Conclusion}

In this work, we proposed DurFlex-EVC, a robust emotional voice conversion framework that supports flexible durations without relying on text or alignment. Exploiting discrete speech units derived from HuBERT, our model encodes content at the unit level and allows flexible duration control by extending unit-level duration prediction to the frame level. We developed a style autoencoder to disentangle emotional style from content, allowing precise manipulation of target styles. Additionally, we introduced a unit aligner to model emotional context at the unit level. A hierarchical stylize encoder further enhances emotional expressiveness by refining style application at both unit and frame levels. To improve synthesis quality, we incorporated a stochastic duration predictor and a diffusion-based generator, ensuring naturalness and diversity in generated speech. Through extensive experiments, DurFlex-EVC demonstrated superior performance over existing models, excelling in both emotional reproduction accuracy and speech quality. The framework effectively generalized to unseen speakers, preserving pronunciation and emotional clarity. These results demonstrate its potential for advancing emotional voice conversion, particularly in applications requiring duration flexibility and expressive emotional transformations.

\bibliographystyle{IEEEtran}
\bibliography{refs}

\begin{thebibliography}{10}
\providecommand{\url}[1]{#1}
\csname url@samestyle\endcsname
\providecommand{\newblock}{\relax}
\providecommand{\bibinfo}[2]{#2}
\providecommand{\BIBentrySTDinterwordspacing}{\spaceskip=0pt\relax}
\providecommand{\BIBentryALTinterwordstretchfactor}{4}
\providecommand{\BIBentryALTinterwordspacing}{\spaceskip=\fontdimen2\font plus
\BIBentryALTinterwordstretchfactor\fontdimen3\font minus \fontdimen4\font\relax}
\providecommand{\BIBforeignlanguage}[2]{{%
\expandafter\ifx\csname l@#1\endcsname\relax
\typeout{** WARNING: IEEEtran.bst: No hyphenation pattern has been}%
\typeout{** loaded for the language `#1'. Using the pattern for}%
\typeout{** the default language instead.}%
\else
\language=\csname l@#1\endcsname
\fi
#2}}
\providecommand{\BIBdecl}{\relax}
\BIBdecl

\bibitem{10065433}
A.~Triantafyllopoulos, B.~W. Schuller, G.~İymen, M.~Sezgin, X.~He, Z.~Yang, P.~Tzirakis, S.~Liu, S.~Mertes, E.~André, R.~Fu, and J.~Tao, ``{An Overview of Affective Speech Synthesis and Conversion in the Deep Learning Era},'' \emph{Proceedings of the IEEE}, vol. 111, no.~10, pp. 1355--1381, 2023.

\bibitem{9829283}
N.~Hussain, E.~Erzin, T.~M. Sezgin, and Y.~Yemez, ``{Training Socially Engaging Robots: Modeling Backchannel Behaviors with Batch Reinforcement Learning},'' \emph{IEEE Trans. Affect. Comput.}, vol.~13, no.~4, pp. 1840--1853, 2022.

\bibitem{8068274}
M.~P. Aylett, A.~Vinciarelli, and M.~Wester, ``{Speech Synthesis for the Generation of Artificial Personality},'' \emph{IEEE Trans. Affect. Comput.}, vol.~11, no.~2, pp. 361--372, 2020.

\bibitem{busso2008iemocap}
C.~Busso, M.~Bulut, C.-C. Lee, A.~Kazemzadeh, E.~Mower, S.~Kim, J.~N. Chang, S.~Lee, and S.~S. Narayanan, ``{IEMOCAP: Interactive emotional dyadic motion capture database},'' \emph{Language resources and evaluation}, vol.~42, pp. 335--359, 2008.

\bibitem{9920692}
J.-H. Jeong, J.-H. Cho, B.-H. Lee, and S.-W. Lee, ``Real-time deep neurolinguistic learning enhances noninvasive neural language decoding for brain–machine interaction,'' \emph{IEEE Trans. on Cybernetics}, vol.~53, no.~12, pp. 7469--7482, 2023.

\bibitem{7160715}
F.~Eyben, K.~R. Scherer, B.~W. Schuller, J.~Sundberg, E.~André, C.~Busso, L.~Y. Devillers, J.~Epps, P.~Laukka, S.~S. Narayanan, and K.~P. Truong, ``{The Geneva Minimalistic Acoustic Parameter Set (GeMAPS) for Voice Research and Affective Computing},'' \emph{IEEE Trans. Affect. Comput.}, vol.~7, no.~2, pp. 190--202, 2016.

\bibitem{5871584}
J.~Sundberg, S.~Patel, E.~Bjorkner, and K.~R. Scherer, ``{Interdependencies among Voice Source Parameters in Emotional Speech},'' \emph{IEEE Trans. Affect. Comput.}, vol.~2, no.~3, pp. 162--174, 2011.

\bibitem{ming16_interspeech}
H.~Ming, D.~Huang, L.~Xie, J.~Wu, M.~Dong, and H.~Li, ``{Deep Bidirectional LSTM Modeling of Timbre and Prosody for Emotional Voice Conversion},'' in \emph{Proc. Interspeech}, 2016.

\bibitem{9687906}
Z.~Du, B.~Sisman, K.~Zhou, and H.~Li, ``{Expressive Voice Conversion: A Joint Framework for Speaker Identity and Emotional Style Transfer},'' in \emph{IEEE Autom. Speech Recognit. Underst. Workshop}, 2021.

\bibitem{aihara2012gmm}
R.~Aihara, R.~Takashima, T.~Takiguchi, and Y.~Ariki, ``{GMM-based emotional voice conversion using spectrum and prosody features},'' \emph{American Journal of Signal Processing}, vol.~2, no.~5, pp. 134--138, 2012.

\bibitem{gao19b_interspeech}
J.~Gao, D.~Chakraborty, H.~Tembine, and O.~Olaleye, ``{Nonparallel Emotional Speech Conversion},'' in \emph{Proc. Interspeech}, 2019.

\bibitem{zhou20d_interspeech}
K.~Zhou, B.~Sisman, M.~Zhang, and H.~Li, ``{Converting Anyone’s Emotion: Towards Speaker-Independent Emotional Voice Conversion},'' in \emph{Proc. Interspeech}, 2020.

\bibitem{9383526}
K.~Zhou, B.~Sisman, and H.~Li, ``{Vaw-Gan For Disentanglement And Recomposition Of Emotional Elements In Speech},'' in \emph{IEEE Spok. Lang. Technol. Workshop}, 2021.

\bibitem{cao20b_interspeech}
Y.~Cao, Z.~Liu, M.~Chen, J.~Ma, S.~Wang, and J.~Xiao, ``{Nonparallel Emotional Speech Conversion Using VAE-GAN},'' in \emph{Proc. Interspeech}, 2020.

\bibitem{9054579}
G.~Rizos, A.~Baird, M.~Elliott, and B.~Schuller, ``{Stargan for Emotional Speech Conversion: Validated by Data Augmentation of End-To-End Emotion Recognition},'' in \emph{IEEE Int. Conf. Acoust., Speech, Signal Process.}, 2020.

\bibitem{Zhu_2017_ICCV}
J.-Y. Zhu, T.~Park, P.~Isola, and A.~A. Efros, ``{Unpaired Image-To-Image Translation Using Cycle-Consistent Adversarial Networks},'' in \emph{Proc. IEEE Int. Conf. Comput. Vis.}, 2017.

\bibitem{Choi_2018_CVPR}
Y.~Choi, M.~Choi, M.~Kim, J.-W. Ha, S.~Kim, and J.~Choo, ``{StarGAN: Unified Generative Adversarial Networks for Multi-Domain Image-to-Image Translation},'' in \emph{Proc. IEEE/CVF Conf. Compt. Vis. Pattern Recognit.}, 2018.

\bibitem{Bao_2017_ICCV}
J.~Bao, D.~Chen, F.~Wen, H.~Li, and G.~Hua, ``{CVAE-GAN: Fine-Grained Image Generation Through Asymmetric Training},'' in \emph{Proc. IEEE Int. Conf. Comput. Vis.}, 2017.

\bibitem{8683865}
C.~Robinson, N.~Obin, and A.~Roebel, ``{Sequence-to-sequence Modelling of F0 for Speech Emotion Conversion},'' in \emph{IEEE Int. Conf. Acoust., Speech, Signal Process.}, 2019.

\bibitem{9053255}
T.-H. Kim, S.~Cho, S.~Choi, S.~Park, and S.-Y. Lee, ``{Emotional Voice Conversion Using Multitask Learning with Text-To-Speech},'' in \emph{IEEE Int. Conf. Acoust., Speech, Signal Process.}, 2020.

\bibitem{zhou21b_interspeech}
K.~Zhou, B.~Sisman, and H.~Li, ``{Limited Data Emotional Voice Conversion Leveraging Text-to-Speech: Two-Stage Sequence-to-Sequence Training},'' in \emph{Proc. Interspeech}, 2021.

\bibitem{9778970}
K.~Zhou, B.~Sisman, R.~Rana, B.~W. Schuller, and H.~Li, ``{Emotion Intensity and its Control for Emotional Voice Conversion},'' \emph{IEEE Trans. Affect. Comput.}, vol.~14, no.~1, pp. 31--48, 2023.

\bibitem{9729483}
S.-H. Lee, H.-R. Noh, W.-J. Nam, and S.-W. Lee, ``{Duration Controllable Voice Conversion via Phoneme-Based Information Bottleneck},'' \emph{IEEE/ACM Trans. Audio, Speech, Lang. Process.}, vol.~30, pp. 1173--1183, 2022.

\bibitem{NEURIPS2019_f63f65b5}
Y.~Ren, Y.~Ruan, X.~Tan, T.~Qin, S.~Zhao, Z.~Zhao, and T.-Y. Liu, ``{FastSpeech: Fast, Robust and Controllable Text to Speech},'' in \emph{Proc. Adv. Neural Inf. Process. Syst.}, 2019.

\bibitem{ren2021fastspeech}
Y.~Ren, C.~Hu, X.~Tan, T.~Qin, S.~Zhao, Z.~Zhao, and T.-Y. Liu, ``{FastSpeech 2: Fast and High-Quality End-to-End Text to Speech},'' in \emph{Proc. Int. Conf. Learn. Repr.}, 2021.

\bibitem{polyak21_interspeech}
A.~Polyak, Y.~Adi, J.~Copet, E.~Kharitonov, K.~Lakhotia, W.-N. Hsu, A.~Mohamed, and E.~Dupoux, ``{Speech Resynthesis from Discrete Disentangled Self-Supervised Representations},'' in \emph{Proc. Interspeech}, 2021.

\bibitem{kreuk-etal-2022-textless}
F.~Kreuk, A.~Polyak, J.~Copet, E.~Kharitonov, T.~A. Nguyen, M.~Rivi{\`e}re, W.-N. Hsu, A.~Mohamed, E.~Dupoux, and Y.~Adi, ``{Textless Speech Emotion Conversion using Discrete {\&} Decomposed Representations},'' in \emph{Proc. Conf. Empir. Methods Nat. Lang. Process}, 2022.

\bibitem{9746484}
B.~van Niekerk, M.-A. Carbonneau, J.~Zaïdi, M.~Baas, H.~Seuté, and H.~Kamper, ``{A Comparison of Discrete and Soft Speech Units for Improved Voice Conversion},'' in \emph{IEEE Int. Conf. Acoust., Speech, Signal Process.}, 2022.

\bibitem{kim23k_interspeech}
H.~Kim, S.~Kim, J.~Yeom, and S.~Yoon, ``{UnitSpeech: Speaker-adaptive Speech Synthesis with Untranscribed Data},'' in \emph{Proc. Interspeech}, 2023.

\bibitem{NEURIPS2020_92d1e1eb}
A.~Baevski, Y.~Zhou, A.~Mohamed, and M.~Auli, ``{wav2vec 2.0: A Framework for Self-Supervised Learning of Speech Representations},'' in \emph{Proc. Adv. Neural Inf. Process. Syst.}, 2020.

\bibitem{Baevski2020vq-wav2vec}
A.~Baevski, S.~Schneider, and M.~Auli, ``vq-{wav2vec: Self-Supervised Learning of Discrete Speech Representations},'' in \emph{Proc. Int. Conf. Learn. Repr.}, 2020.

\bibitem{babu2021xls}
A.~Babu, C.~Wang, A.~Tjandra, K.~Lakhotia, Q.~Xu, N.~Goyal, K.~Singh, P.~von Platen, Y.~Saraf, J.~Pino \emph{et~al.}, ``{XLS-R: Self-supervised cross-lingual speech representation learning at scale},'' \emph{arXiv preprint arXiv:2111.09296}, 2021.

\bibitem{9585401}
W.-N. Hsu, B.~Bolte, Y.-H.~H. Tsai, K.~Lakhotia, R.~Salakhutdinov, and A.~Mohamed, ``{HuBERT: Self-Supervised Speech Representation Learning by Masked Prediction of Hidden Units},'' \emph{IEEE/ACM Trans. Audio, Speech, Lang. Process.}, vol.~29, pp. 3451--3460, 2021.

\bibitem{pmlr-v162-qian22b}
K.~Qian, Y.~Zhang, H.~Gao, J.~Ni, C.-I. Lai, D.~Cox, M.~Hasegawa-Johnson, and S.~Chang, ``{ContentVec: An Improved Self-Supervised Speech Representation by Disentangling Speakers},'' in \emph{Proc. Int. Conf. on Mach. Learn.}, 2022.

\bibitem{lee23i_interspeech}
S.-H. Lee, H.-Y. Choi, H.-S. Oh, and S.-W. Lee, ``{HierVST: Hierarchical Adaptive Zero-shot Voice Style Transfer},'' in \emph{Proc. Interspeech}, 2023.

\bibitem{9747814}
Z.~Chen, S.~Chen, Y.~Wu, Y.~Qian, C.~Wang, S.~Liu, Y.~Qian, and M.~Zeng, ``{Large-Scale Self-Supervised Speech Representation Learning for Automatic Speaker Verification},'' in \emph{IEEE Int. Conf. Acoust., Speech, Signal Process.}, 2022.

\bibitem{10089511}
J.~Wagner, A.~Triantafyllopoulos, H.~Wierstorf, M.~Schmitt, F.~Burkhardt, F.~Eyben, and B.~W. Schuller, ``{Dawn of the Transformer Era in Speech Emotion Recognition: Closing the Valence Gap},'' \emph{IEEE Trans. Pattern Anal. Mach. Intell.}, vol.~45, no.~9, pp. 10\,745--10\,759, 2023.

\bibitem{9625818}
N.~Zeghidour, A.~Luebs, A.~Omran, J.~Skoglund, and M.~Tagliasacchi, ``{SoundStream: An End-to-End Neural Audio Codec},'' \emph{IEEE/ACM Trans. Audio, Speech, Lang. Process.}, vol.~30, pp. 495--507, 2022.

\bibitem{dfossez2023high}
A.~D{\'e}fossez, J.~Copet, G.~Synnaeve, and Y.~Adi, ``{High Fidelity Neural Audio Compression},'' \emph{Transactions on Machine Learning Research}, 2023, featured Certification, Reproducibility Certification.

\bibitem{yang2023uniaudio}
D.~Yang, J.~Tian, X.~Tan, R.~Huang, S.~Liu, X.~Chang, J.~Shi, S.~Zhao, J.~Bian, X.~Wu \emph{et~al.}, ``{UniAudio: An Audio Foundation Model Toward Universal Audio Generation},'' \emph{arXiv preprint arXiv:2310.00704}, 2023.

\bibitem{NEURIPS2020_5c3b99e8}
J.~Kim, S.~Kim, J.~Kong, and S.~Yoon, ``{Glow-TTS: A Generative Flow for Text-to-Speech via Monotonic Alignment Search},'' in \emph{Proc. Adv. Neural Inf. Process. Syst.}, 2020.

\bibitem{pmlr-v139-popov21a}
V.~Popov, I.~Vovk, V.~Gogoryan, T.~Sadekova, and M.~Kudinov, ``{Grad-TTS: A Diffusion Probabilistic Model for Text-to-Speech},'' in \emph{Proc. Int. Conf. on Mach. Learn.}, 2021.

\bibitem{mcauliffe17_interspeech}
M.~McAuliffe, M.~Socolof, S.~Mihuc, M.~Wagner, and M.~Sonderegger, ``{Montreal Forced Aligner: Trainable Text-Speech Alignment Using Kaldi},'' in \emph{Proc. Interspeech}, 2017.

\bibitem{8639535}
H.~Kameoka, T.~Kaneko, K.~Tanaka, and N.~Hojo, ``Stargan-vc: non-parallel many-to-many voice conversion using star generative adversarial networks,'' in \emph{IEEE Spok. Lang. Technol. Workshop}, 2018.

\bibitem{pmlr-v97-qian19c}
K.~Qian, Y.~Zhang, S.~Chang, X.~Yang, and M.~Hasegawa-Johnson, ``{A}uto{VC}: Zero-shot voice style transfer with only autoencoder loss,'' in \emph{Proc. Int. Conf. on Mach. Learn.}, 2019, pp. 5210--5219.

\bibitem{8683282}
K.~Tanaka, H.~Kameoka, T.~Kaneko, and N.~Hojo, ``{ATTS2S-VC: Sequence-to-sequence Voice Conversion with Attention and Context Preservation Mechanisms},'' in \emph{IEEE Int. Conf. Acoust., Speech, Signal Process.}, 2019.

\bibitem{9413973}
T.~Hayashi, W.-C. Huang, K.~Kobayashi, and T.~Toda, ``Non-autoregressive sequence-to-sequence voice conversion,'' in \emph{IEEE Int. Conf. Acoust., Speech, Signal Process.}, 2021.

\bibitem{NEURIPS2020_c5d73680}
J.~Kong, J.~Kim, and J.~Bae, ``Hifi-gan: Generative adversarial networks for efficient and high fidelity speech synthesis,'' in \emph{Proc. Adv. Neural Inf. Process. Syst.}, 2020.

\bibitem{8683143}
R.~Prenger, R.~Valle, and B.~Catanzaro, ``Waveglow: A flow-based generative network for speech synthesis,'' in \emph{Proc. Adv. Neural Inf. Process. Syst.}, 2019.

\bibitem{kong2021diffwave}
Z.~Kong, W.~Ping, J.~Huang, K.~Zhao, and B.~Catanzaro, ``Diffwave: A versatile diffusion model for audio synthesis,'' in \emph{Proc. Int. Conf. Learn. Repr.}, 2021.

\bibitem{8461368}
J.~Shen, R.~Pang, R.~J. Weiss, M.~Schuster, N.~Jaitly, Z.~Yang, Z.~Chen, Y.~Zhang, Y.~Wang, R.~Skerrv-Ryan, R.~A. Saurous, Y.~Agiomvrgiannakis, and Y.~Wu, ``{Natural TTS Synthesis by Conditioning Wavenet on MEL Spectrogram Predictions},'' in \emph{IEEE Int. Conf. Acoust., Speech, Signal Process.}, 2018.

\bibitem{Li_Liu_Liu_Zhao_Liu_2019}
N.~Li, S.~Liu, Y.~Liu, S.~Zhao, and M.~Liu, ``{Neural Speech Synthesis with Transformer Network},'' \emph{Proc. AAAI Conf. Artif. Intell.}, vol.~33, no.~01, 2019.

\bibitem{8639647}
C.-c. Yeh, P.-c. Hsu, J.-c. Chou, H.-y. Lee, and L.-s. Lee, ``{Rhythm-Flexible Voice Conversion Without Parallel Data Using Cycle-GAN Over Phoneme Posteriorgram Sequences},'' in \emph{IEEE Spok. Lang. Technol. Workshop}, 2018.

\bibitem{yang22t_interspeech}
Z.~Yang, X.~Jing, A.~Triantafyllopoulos, M.~Song, I.~Aslan, and B.~W. Schuller, ``{An Overview \& Analysis of Sequence-to-Sequence Emotional Voice Conversion},'' in \emph{Proc. Interspeech}, 2022.

\bibitem{NEURIPS2022_4730d10b}
R.~Huang, Y.~Ren, J.~Liu, C.~Cui, and Z.~Zhao, ``{GenerSpeech: Towards Style Transfer for Generalizable Out-Of-Domain Text-to-Speech},'' in \emph{Proc. Adv. Neural Inf. Process. Syst.}, 2022.

\bibitem{chen2021adaspeech}
M.~Chen, X.~Tan, B.~Li, Y.~Liu, T.~Qin, sheng zhao, and T.-Y. Liu, ``{AdaSpeech: Adaptive Text to Speech for Custom Voice},'' in \emph{Proc. Int. Conf. Learn. Repr.}, 2021.

\bibitem{10095515}
M.~Kang, D.~Min, and S.~J. Hwang, ``{Grad-StyleSpeech: Any-Speaker Adaptive Text-to-Speech Synthesis with Diffusion Models},'' in \emph{IEEE Int. Conf. Acoust., Speech, Signal Process.}, 2023.

\bibitem{pmlr-v139-kim21f}
J.~Kim, J.~Kong, and J.~Son, ``{Conditional Variational Autoencoder with Adversarial Learning for End-to-End Text-to-Speech},'' in \emph{Proc. Int. Conf. on Mach. Learn.}, 2021.

\bibitem{ZHOU20221}
K.~Zhou, B.~Sisman, R.~Liu, and H.~Li, ``{Emotional voice conversion: Theory, databases and ESD},'' \emph{Speech Communication}, vol. 137, pp. 1--18, 2022.

\bibitem{lee2023bigvgan}
S.~gil Lee, W.~Ping, B.~Ginsburg, B.~Catanzaro, and S.~Yoon, ``{BigVGAN: A Universal Neural Vocoder with Large-Scale Training},'' in \emph{The Eleventh Proc. Int. Conf. Learn. Repr.}, 2023.

\bibitem{zen19_interspeech}
H.~Zen, V.~Dang, R.~Clark, Y.~Zhang, R.~J. Weiss, Y.~Jia, Z.~Chen, and Y.~Wu, ``{LibriTTS: A Corpus Derived from LibriSpeech for Text-to-Speech},'' in \emph{Proc. Interspeech}, 2019.

\bibitem{saeki22c_interspeech}
T.~Saeki, D.~Xin, W.~Nakata, T.~Koriyama, S.~Takamichi, and H.~Saruwatari, ``{UTMOS: UTokyo-SaruLab System for VoiceMOS Challenge 2022},'' in \emph{Proc. Interspeech}, 2022.

\bibitem{xu22b_interspeech}
Q.~Xu, A.~Baevski, and M.~Auli, ``{Simple and Effective Zero-shot Cross-lingual Phoneme Recognition},'' in \emph{Proc. Interspeech}, 2022.

\bibitem{pmlr-v202-radford23a}
A.~Radford, J.~W. Kim, T.~Xu, G.~Brockman, C.~Mcleavey, and I.~Sutskever, ``{Robust Speech Recognition via Large-Scale Weak Supervision},'' in \emph{Proc. Int. Conf. on Mach. Learn.}, 2023.

\bibitem{ma2023emotion2vec}
Z.~Ma, Z.~Zheng, J.~Ye, J.~Li, Z.~Gao, S.~Zhang, and X.~Chen, ``emotion2vec: Self-supervised pre-training for speech emotion representation,'' \emph{Proc. ACL Findings}, 2024.

\bibitem{10003644}
K.~Zhou, B.~Sisman, R.~Rana, B.~W. Schuller, and H.~Li, ``{Speech Synthesis with Mixed Emotions},'' \emph{IEEE Trans. Affect. Comput.}, pp. 1--16, 2022.

\bibitem{wang21h_interspeech}
J.~Wang, J.~Li, X.~Zhao, Z.~Wu, S.~Kang, and H.~Meng, ``Adversarially learning disentangled speech representations for robust multi-factor voice conversion,'' in \emph{Interspeech 2021}, 2021, pp. 846--850.

\bibitem{9747987}
T.~Li, X.~Wang, Q.~Xie, Z.~Wang, and L.~Xie, ``Cross-speaker emotion disentangling and transfer for end-to-end speech synthesis,'' \emph{IEEE/ACM Transactions on Audio, Speech, and Language Processing}, vol.~30, pp. 1448--1460, 2022.

\bibitem{10106091}
G.~Zhang, Y.~Qin, W.~Zhang, J.~Wu, M.~Li, Y.~Gai, F.~Jiang, and T.~Lee, ``iemotts: Toward robust cross-speaker emotion transfer and control for speech synthesis based on disentanglement between prosody and timbre,'' \emph{IEEE/ACM Transactions on Audio, Speech, and Language Processing}, vol.~31, pp. 1693--1705, 2023.

\bibitem{9814838}
S.~Chen, C.~Wang, Z.~Chen, Y.~Wu, S.~Liu, Z.~Chen, J.~Li, N.~Kanda, T.~Yoshioka, X.~Xiao, J.~Wu, L.~Zhou, S.~Ren, Y.~Qian, Y.~Qian, J.~Wu, M.~Zeng, X.~Yu, and F.~Wei, ``{WavLM: Large-Scale Self-Supervised Pre-Training for Full Stack Speech Processing},'' \emph{IEEE J. Sel. Top. Signal Process.}, vol.~16, no.~6, pp. 1505--1518, 2022.

\bibitem{NEURIPS2022_69c754f5}
S.-H. Lee, S.-B. Kim, J.-H. Lee, E.~Song, M.-J. Hwang, and S.-W. Lee, ``{HierSpeech: Bridging the Gap between Text and Speech by Hierarchical Variational Inference using Self-supervised Representations for Speech Synthesis},'' in \emph{Proc. Adv. Neural Inf. Process. Syst.}, 2022.

\bibitem{10.3115/1073083.1073135}
K.~Papineni, S.~Roukos, T.~Ward, and W.-J. Zhu, ``{BLEU: A Method for Automatic Evaluation of Machine Translation},'' in \emph{Proceedings of the 40th Annual Meeting on Association for Computational Linguistics}, ser. ACL '02.\hskip 1em plus 0.5em minus 0.4em\relax USA: Association for Computational Linguistics, 2002, p. 311–318.

\bibitem{pmlr-v139-min21b}
D.~Min, D.~B. Lee, E.~Yang, and S.~J. Hwang, ``Meta-stylespeech : Multi-speaker adaptive text-to-speech generation,'' in \emph{Proc. Int. Conf. on Mach. Learn.}, 2021.

\bibitem{pmlr-v37-ganin15}
Y.~Ganin and V.~Lempitsky, ``{Unsupervised Domain Adaptation by Backpropagation},'' in \emph{Proc. Int. Conf. on Mach. Learn.}, 2015.

\bibitem{8341805}
M.~Tahon, G.~Lecorvé, and D.~Lolive, ``{Can We Generate Emotional Pronunciations for Expressive Speech Synthesis?}'' \emph{IEEE Trans. Affect. Comput.}, vol.~11, no.~4, pp. 684--695, 2020.

\bibitem{xiao2022tackling}
Z.~Xiao, K.~Kreis, and A.~Vahdat, ``{Tackling the Generative Learning Trilemma with Denoising Diffusion GANs},'' in \emph{Proc. Int. Conf. Learn. Repr.}, 2022.

\bibitem{10517426}
H.-S. Oh, S.-H. Lee, and S.-W. Lee, ``Diffprosody: Diffusion-based latent prosody generation for expressive speech synthesis with prosody conditional adversarial training,'' \emph{IEEE/ACM Trans. Audio, Speech, Lang. Process.}, vol.~32, pp. 2654--2666, 2024.

\bibitem{Lee_Yoon_Noh_Kim_Lee_2021}
S.-H. Lee, H.-W. Yoon, H.-R. Noh, J.-H. Kim, and S.-W. Lee, ``Multi-spectrogan: High-diversity and high-fidelity spectrogram generation with adversarial style combination for speech synthesis,'' \emph{Proc. AAAI Conf. Artif. Intell.}, 2021.

\bibitem{NEURIPS2021_0266e33d}
S.-H. Lee, J.-H. Kim, H.~Chung, and S.-W. Lee, ``Voicemixer: Adversarial voice style mixup,'' in \emph{Proc. Adv. Neural Inf. Process. Syst.}, 2021.

\bibitem{10381805}
S.-B. Kim, S.-H. Lee, H.-Y. Choi, and S.-W. Lee, ``Audio super-resolution with robust speech representation learning of masked autoencoder,'' \emph{IEEE/ACM Trans. Audio, Speech, Lang. Process.}, vol.~32, pp. 1012--1022, 2024.

\bibitem{10447331}
S.-B. Kim, S.-H. Lee, and S.-W. Lee, ``Transentence: speech-to-speech translation via language-agnostic sentence-level speech encoding without language-parallel data,'' in \emph{IEEE Int. Conf. Acoust., Speech, Signal Process.}, 2024, pp. 12\,722--12\,726.

\bibitem{9003829}
X.~Zhu, S.~Yang, G.~Yang, and L.~Xie, ``{Controlling Emotion Strength with Relative Attribute for End-to-End Speech Synthesis},'' in \emph{IEEE Autom. Speech Recognit. Underst. Workshop}, 2019.

\bibitem{9383524}
Y.~Lei, S.~Yang, and L.~Xie, ``{Fine-Grained Emotion Strength Transfer, Control and Prediction for Emotional Speech Synthesis},'' in \emph{IEEE Spok. Lang. Technol. Workshop}, 2021.

\bibitem{9053732}
S.-Y. Um, S.~Oh, K.~Byun, I.~Jang, C.~Ahn, and H.-G. Kang, ``{Emotional Speech Synthesis with Rich and Granularized Control},'' in \emph{IEEE Int. Conf. Acoust., Speech, Signal Process.}, 2020.

\bibitem{9747098}
C.-B. Im, S.-H. Lee, S.-B. Kim, and S.-W. Lee, ``{EMOQ-TTS: Emotion Intensity Quantization for Fine-Grained Controllable Emotional Text-to-Speech},'' in \emph{IEEE Int. Conf. Acoust., Speech, Signal Process.}, 2022.

\bibitem{10447981}
D.-M. Byun, S.-H. Lee, J.-S. Hwang, and S.-W. Lee, ``Midi-voice: Expressive zero-shot singing voice synthesis via midi-driven priors,'' in \emph{IEEE Int. Conf. Acoust., Speech, Signal Process.}, 2024, pp. 12\,622--12\,626.

\end{thebibliography}

\newpage

\begin{IEEEbiography}[
{\includegraphics[width=1in,height=1.25in,clip,keepaspectratio]{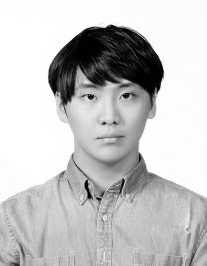}}]{Hyung-Seok~Oh}
received the B.S. degree in Computer Science and Engineering from Konkuk University, Seoul, South Korea, in 2021. He is currently working toward an integrated master's and Ph.D. degree with the Department of Artificial Intelligence, Korea University, Seoul, South Korea. His research interests include artificial intelligence and audio signal processing.
\end{IEEEbiography}
\vskip -2\baselineskip plus -1fil
\begin{IEEEbiography}[{\includegraphics[width=1in,height=1.25in,clip,keepaspectratio]{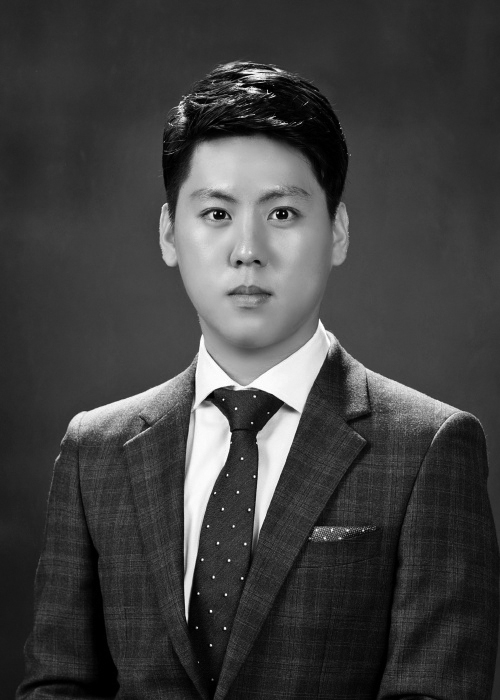}}]{Sang-Hoon~Lee}
received the B.S. degree in life science from Dongguk University, Seoul, South Korea, in 2016 and received the Ph.D. degree in Brain and Cognitive Engineering from Korea University, Seoul, South Korea, in 2023. He was a postdoctoral researcher in AI Research Center, Korea University, Seoul, South Korea, in 2024. He is currently an assistant professor in Department of Software and Computer Engineering from Ajou University, Suwon, South Korea. His current research interests include artificial intelligence and audio signal processing.
\end{IEEEbiography}
\vskip -2\baselineskip plus -1fil
\begin{IEEEbiography}[{\includegraphics[width=1in,height=1.25in,clip,keepaspectratio]{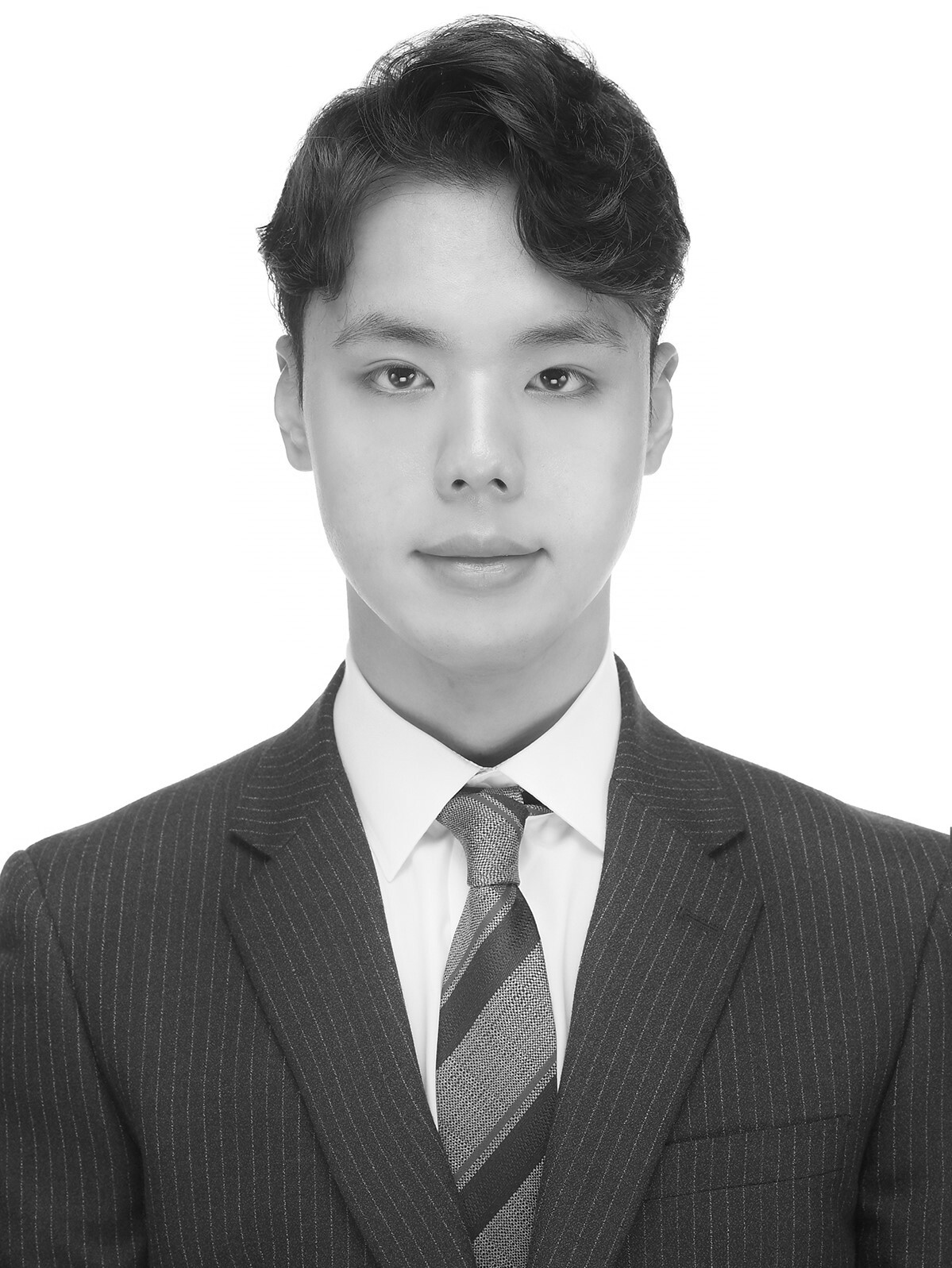}}]{Deok-Hyeon~Cho}
received the B.S. degree in Applied Mathematics from Hanyang University ERICA Campus, Ansan, South Korea, in 2022. He is currently working toward an integrated master's and Ph.D. degree with the Department of Artificial Intelligence, Korea University, Seoul, South Korea. His research interests include artificial intelligence and audio signal processing.
\end{IEEEbiography}
\vskip -2\baselineskip plus -1fil
\begin{IEEEbiography}[{\includegraphics[width=1in,height=1.25in,clip,keepaspectratio]{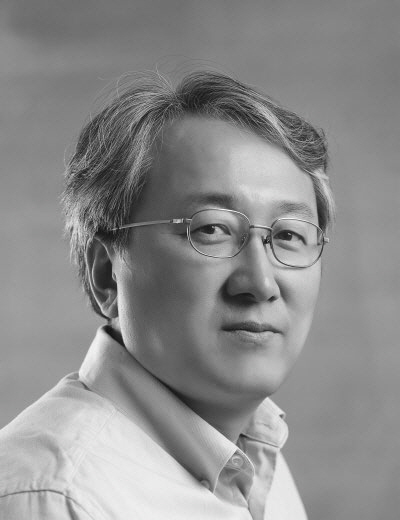}}]{Seong-Whan~Lee}
(Fellow, IEEE) received the B.S. degree in computer science and statistics from Seoul National University, South Korea, in 1984, and the M.S. and Ph.D. degrees in computer science from the Korea Advanced Institute of Science and Technology, South Korea, in 1986 and 1989, respectively. He is currently the Head of the Department of Artificial Intelligence, Korea University, Seoul. His current research interests include artificial intelligence, pattern recognition, and brain engineering. He is a Fellow of the International Association of Pattern Recognition (IAPR), the Korea Academy of Science and Technology, and the National Academy of Engineering of Korea.
\end{IEEEbiography}

\end{document}